# Conductivity of twin walls - surface junctions in ferroelastics: interplay of deformation potential, octahedral rotations, improper ferroelectricity and flexoelectric coupling

Eugene A. Eliseev[1], Anna N. Morozovska[2*], Yijia Gu[3], Albina Y. Borisevich[4],
Long-Qing Chen[3] and Venkatraman Gopalan[3], and Sergei V. Kalinin[4]

[1] Institute for Problems of Materials Science, National Academy of Science of Ukraine,
3, Krjijanovskogo, 03142 Kiev, Ukraine

[2] Institute of Physics, National Academy of Science of Ukraine, 46, pr. Nauki, 03028 Kiev, Ukraine

[3] Department of Materials Science and Engineering, Pennsylvania State University,
University Park, Pennsylvania 16802, USA

[4] Center for Nanophase Materials Science,
Oak Ridge National Laboratory, Oak Ridge, TN, 37831

Electronic and structural phenomena at the twin domain wall-surface junctions in the ferroelastic materials are analyzed. Carriers accumulation caused by the strain-induced band structure changes originated via the deformation potential mechanism, structural order parameter gradient, rotostriction and flexoelectric coupling is explored. Approximate analytical results show that inhomogeneous elastic strains, which exist in the vicinity of the twin walls – surface junctions due to the rotostriction coupling, decrease the local band gap via the deformation potential and flexoelectric coupling mechanisms. This is the *direct mechanism* of the twin walls static conductivity in ferroelastics and, by extension, in multiferroics and ferroelectrics. On the other hand, flexoelectric and rotostriction coupling leads to the appearance of the improper polarization and electric fields proportional to the structural order parameter gradient in the vicinity of the twin walls – surface junctions. The "flexo-roto" fields leading to the carrier accumulation are considered as *indirect mechanism* of the twin walls conductivity. Comparison of the direct and indirect mechanisms illustrates complex range of phenomena directly responsible for domain walls static conductivity in materials with multiple order parameters.

---

*Corresponding author e-mail morozo@i.com.ua

## 1. Introduction

The interactions between the soft-phonon driven lattice instabilities and electronic phenomena have fascinated physicists for more than half a century. Traditionally, the relevant areas included domain instabilities and light-induced phenomena in ferroelectrics [1, 2, 3, 4, 5], electronic phenomena at the domain walls and surfaces [6, 7]. The first set of phenomena specifically addresses the interaction of ferroelectric order parameter fields with non-equilibrium charge carriers and has also found renewed interest due to ferroelectric photovoltaics [8, 9]. The second includes electronic [2, 10, 11, 12] and now electrochemical [13, 14, 15, 16, 17, 18, 19] phenomena induced by ferroelectric polarization charge at surfaces and interfaces. From this perspective, ferroic walls offer arguably the simplest system for exploration of the interplay between ferroic and electronic phenomena due to continuity of atomic lattice and minimal contribution of chemical and electrochemical effects.

The earliest theoretical predictions of domain walls conductivity in ferroelectric-semiconductors was made by Guro et al. [20] in 1969, the mechanism stemming from the compensation of polarization charge discontinuity by mobile carriers in the material. This model was further developed for uniaxial [21, 22] and multiaxial ferroelectrics [23], and improper ferroelectrics [24]. Numerous experimental justifications appeared after the development of scanning probe microscopy (SPM) techniques, capable of probing the conductance on the nanoscale, in multiferroic $BiFeO_3$ [25, 26, 27, 28], ferroelectrics $Pb(Zr,Ti)O_3$ [29, 30], $ErMnO_3$ [31] and $LiNbO_3$ doped with MgO [32]. Interestingly, the preponderance of the experimental results [25-28, 30] report about the conductivity of the nominally uncharged domain walls in multiferroics. Some of the recent studies demonstrate hysteretic conductance of domain walls and presence of multiple remnant conduction states [26, 30, 33] behaviour ascribed to the role of metastable tilted wall configurations pinned by the structural defects. The recent report [30] on metallic conductivity of domain walls in $Pb(Zr,Ti)O_3$ provides strong evidence towards the semiconductor-ferroelectric model of charged domain walls.

However, the vast majority of materials that is now being explored in the context of domain wall mediated electronic phenomena possesses multiple structural instabilities, and is often incipient or proper ferroelectric and ferroelastic. From this context, it is germane to consider the progression of theoretical models for wall structures. On the most basic level, thermo-dynamical studies [21, 22 and 23] consider the carrier accumulation by the strongly charged perpendicular (or “counter”) and inclined 180-degree ferroelectric domain walls respectively. Depending on the incline angle, the incline walls can be strongly charged, weakly charged or uncharged [23]. Free carrier concentration, band bending and enhanced electromechanical response at charged 90-degree domain walls in $BaTiO_3$ was recently considered in Ref. [34]. Fiebig et al [31] demonstrated that the electrical conductance at the interfacial ferroelectric domain walls in hexagonal $ErMnO_3$ (and in analogous material $YMnO_3$) is a continuous function of the domain wall orientation, with a range of an order of magnitude variation between head-to-head and tail-to-tail domains walls. The variation is the combined consequence of carrier accumulation and band-structure changes at the walls. So, the origin of the charged domain walls conductivity seems clear enough: the bound charge variation across the wall causes the electric field that in turn attracts screening carriers and causes band-structure changes.

However, the structural and electronic phenomena at the nominally uncharged domain walls in ferroelectrics-ferroelastics cannot be covered by aforementioned studies [20-23, 31, 34]. Recently, it was proposed [30, 23] that the flexoelectric coupling can lead to the appearance of the inhomogeneous electric fields proportional to the polarization gradient across the nominally uncharged domain wall (named as flexoelectric field [30, 23]) and to the field proportional to the structural order parameter gradient (named as roto-flexo field [35, 36]). Notably, strain gradients are expected to induce polarization near the surfaces and interfaces via the flexoelectric effect [37, 38] in all materials, since they are flexoelectrics [39, 40, 41]. Roto-flexo fields can then exist in wide class of materials with oxygen octahedra rotations [42].

We further note that additional factor affecting the domain wall behavior is the strain and field-driven segregation of mobile ions. In particular Salje et al [43] have analyzed the surface structure of domain twin walls in ferroelastics using molecular dynamics simulation. Using analytical models Rychetsky [44] considered the deformation of crystal surfaces in ferroelastic materials caused by antiphase domain boundaries. Using empirical force fields Salje and Lee [45] numerically studied the interaction of oxygen vacancies in the ferroelastic $CaTiO_3$ [100] twin walls. Note, that improper ferroelectricity induced by octahedral rotations exists in $YMnO_3$ [46], $Ca_3Mn_2O_7$ [47], $CaTiO_3$ [48] and their interfaces [49]. In particular Salje et al directly observed ferrielectric polarization at ferroelastic domain boundaries in $CaTiO_3$ by aberration-corrected Transmission Electron Microscopy (TEM) at room temperature [50].

We further note that exploration of domain walls properties by Scanning Probe Microscopy (SPM) necessarily involves probing not only the wall properties per se, but also the responses of the wall – surface junction since the latter is the inherent part of conduction path. In comparison, wall-back electrode junction is more distributed allowing for smaller resistances and higher contribution of defect-mediated conduction paths, and hence its properties are relatively less important. For the case of wall-surface junction, the relaxation of elastic stresses will lead to electric and/or elastic fields with power decay [51, 52]. Even for classical ferroelectrics, the carriers accumulation caused by the wall - surface junctions is studied only recently. [53]. These considerations motivate us to study analytically free carriers accumulation caused by the twin wall - surface junctions in ferroelastics - incipient ferroelectrics, and explore the role of improper ferroelectricity induced by the inhomogeneous octahedral rotations.

In this manuscript, we analytically solve the 2D problem of wall-surface junction for the model $CaTiO_3$ material. The paper is organized as following. Elastic fields caused by the twin wall – surface junction is calculated analytically and analyzed in the Section 2. In the Section 3 we consider the improper ferroelectricity appeared at the twin wall – surface junction. Section 4 is devoted to the carriers accumulation at the twin wall – surface junction. Vacancies segregation at the twin wall – surface junction is estimated in the Section 5. The relevant mathematical details and materials parameters are provided in the Supplemental materials.

## 2. Elastic fields caused by the twin wall – surface junction

Here, we analyze the structure of the ***elastic fields*** created by the domain wall – surface junction in ferroelastics using the perturbation method proposed by Rychetsky [44]. In the first approximation, the surface displacement $u_i^S(\mathbf{x})$ at location **x** induced by the elastic wall – surface junction is given by the convolution of the corresponding Green function with the elastic stress field, $\sigma_{jk}^0$, unperturbed by the surface influence:

$$u_i^S(\mathbf{x}) = \int_{-\infty}^{\infty} d\xi_1 \int_{-\infty}^{\infty} d\xi_2 G_{ij}(x_1-\xi_1, x_2-\xi_2, x_3)\sigma_{jk}^0(\xi_1,\xi_2)n_k \qquad (1)$$

Corresponding Green's tensor $G_{ij}(\mathbf{x}-\xi)$ for elastically isotropic half-space is given by Lur'e [54] and Landau and Lifshitz [55] (see **Appendix A**); $n_k$ is the outer normal to the mechanically free surface $x_3 = 0$. Geometry of calculations is shown in **Fig.1a**. Hereinafter we consider the semi-infinite mechanically free crystal, but not the film on the substrate. However, the approach can be extended to the film case if one will use the Green function corresponding to the elastic problem of mechanically clamped/free film.

Inhomogeneous elastic stresses $\sigma_{jk}^0(\xi_1,\xi_2)$ originate from the rotostriction coupling with the structural order parameter variation appeared in the vicinity of ferroelastic 90-degree twins. The order parameter describing oxygen octahedral rotations can be chosen either as the rotation angle or the displacement of as an appropriate oxygen atom from its cubic position [56, 57, 58]. The behavior of tetragonal ferroelastics (e.g. $SrTiO_3$ below 105 K) in the low-symmetry phase can be described by a single axial vector $\mathbf{\Phi} = (\Phi_1, \Phi_2, \Phi_3)$ [35, 36, 58]; while description of orthorhombic ferroelastics (e.g. $CaTiO_3$ at room temperature) requires two axial vectors $\mathbf{\Phi} = (\Phi_1, \Phi_2, \Phi_3)$ and $\mathbf{\Psi} = (\Psi_1, \Psi_2, \Psi_3)$ [59].

Here, for tetragonal ferroelastics we consider 90-degree twins with nonzero oxygen displacement components $\Phi_1$ and $\Phi_2$ [56]. For orthorhombic ferroelastics we consider 90-degree twins nonzero oxygen displacement nonzero components are $\Phi_1$, $\Phi_2$ and $\Psi_3$ (see **Fig.1b**) [60]. In order to obtain approximate analytical expressions for $\sigma^0_{jk}$ unperturbed by the surface in orthorhombic ferroelastics, approximate analytical expressions for the structural order parameter components were used as: $\Phi_1 \approx \Phi_S\left(1 + a\cosh^{-2}(x_1/w)\right)$, $\Phi_2 \approx \Phi_S \tanh(x_1/w)$, $\Psi_3 \approx \Psi_S\left(1 + b\cosh^{-2}(x_1/w)\right)$, where $\Phi_S$ and $\Psi_S$ are spontaneous values, $w$ is the intrinsic width of the twin wall in the bulk; and the amplitudes $a$ and $b$ appeared much smaller than unity. Rigorously speaking the wall width can be obtained from DFT calculations [61] or STEM measurements [50], and the profile functions are consistent with GLD theory and experiments. In **Appendix B** we list analytical expressions for several cases of the elastic stresses $\sigma^0_{jk}$, namely typical 90-degree twins in ferroelastics with tetragonal and orthorhombic space symmetry.

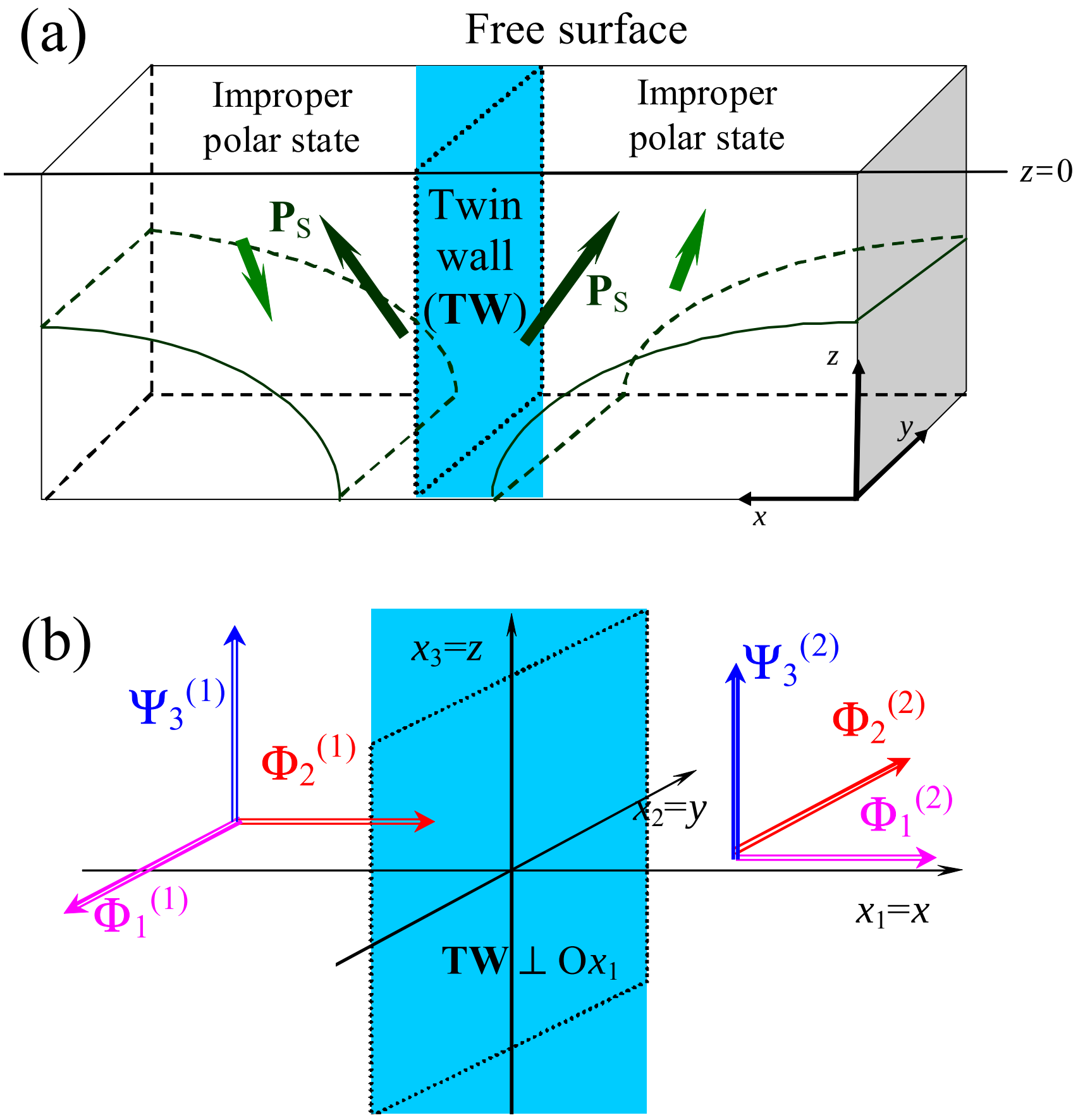

**Figure 1.** (Color online) (a) Twin wall (TW) near the film surface. (b) Orientation of order parameters of the domains and TW with respect to pseudo-cubic axes $Ox_1$, $Ox_2$ and $Ox_3$ for the case of head-to-tail TW. For orthorhombic ferroelastics we consider 90-degree twins nonzero oxygen displacement nonzero components are $\Phi_1$, $\Phi_2$ and $\Psi_3$. TW should be perpendicular to $[100]$ or $[010]$ directions. Coordinates $x_1 = x$, $x_2 = y$ and $x_3 = z$.

The strains $u_{kl}(x,z)$ can be calculated from Eq.(1) using the perturbation approach as $u_{kl}(x,z)=u^0_{kl}+\frac{1}{2}\left(\frac{\partial u^S_i}{\partial x_j}+\frac{\partial u^S_j}{\partial x_i}\right)$, where $u^0_{kl}$ is the strain field of the twin unperturbed by the surface. The stresses $\sigma_{kl}(x,z)$ are listed in **Appendix B**. After lengthy calculations we obtained Pade approximations [62] for nonzero strains:

$$u_{xx}(x,z)\approx u^0_{xx}+\frac{(1+\nu)\,w\cdot\delta\sigma}{Y\left(x^2+(w+z)^2\right)^3}\begin{pmatrix}-\left(w^2+x^2\right)(w+z)^3-x^2z\left(3w(w+z)+x^2\right)+\\ \left(x^2+(w+z)^2\right)\left(z\left(x^2+(w+z)^2\right)+2w(w+z)^2\right)\nu\end{pmatrix}, \tag{2a}$$

$$u_{xz}(x,z)\approx-\frac{(1+\nu)\,\delta\sigma wxz\left(4w(w+z)^2+z\left(x^2+(w+z)^2\right)\right)}{Y\left(x^2+(w+z)^2\right)^3},\qquad u_{yy}(x,z)\approx u^0_{yy}, \tag{2b}$$

$$u_{zz}(x,z)\approx u^0_{zz}+\frac{(1+\nu)w\cdot\delta\sigma}{Y\left(x^2+(w+z)^2\right)^3}\begin{pmatrix}-\left(w^2+3wz+z^2\right)(w+z)^3-x^2\left(w^3+z^3\right)+\\ \left(x^2+(w+z)^2\right)\left(z\left(x^2+(w+z)^2\right)+2w(w+z)^2\right)\nu\end{pmatrix}. \tag{2c}$$

Here, $\nu$ is Poisson ratio, $Y$ is the Young modulus, $w$ is the intrinsic half-width of the twin wall in the bulk, coordinates $x_1=x$, $x_2=y$ and $x_3=z$. The stress $\delta\sigma$ for the twins in tetragonal ferroelastics is listed in **Appendix B.** The stress $\delta\sigma$ for the twins in orthorhombic ferroelastics:

$$\delta\sigma\approx\frac{s_{11}\left(R_{12}+(V_{122}+6V_{112})\Phi_S^2\right)-s_{12}\left(R_{11}+(V_{111}+6V_{112})\Phi_S^2\right)}{s_{11}^2-s_{12}^2}\Phi_S^2. \tag{3a}$$

The strain field of the twin unperturbed by the surface is:

$$u^0_{xx}\approx R_{11}\Phi_S^2+R_{12}\Phi_2^2+V_{111}\Phi_S^4+6V_{112}\Phi_S^2\Phi_2^2+V_{122}\Phi_2^4+W_{122}\Psi_S^4+\\ \frac{s_{12}\left(\Phi_S^2-\Phi_2^2\right)}{s_{11}+s_{12}}\left(R_{11}+R_{12}+(V_{111}+V_{122})\left(\Phi_S^2+\Phi_2^2\right)+6(V_{112}+V_{123})\Phi_S^2\right), \tag{3b}$$

$$u^0_{yy}\approx(R_{11}+R_{12})\Phi_S^2+(V_{122}+6V_{112}+V_{111})\Phi_S^4+W_{122}\Psi_S^4, \tag{3c}$$

$$u^0_{zz}\approx 2R_{12}\Phi_S^2+(2V_{122}+6V_{123})\Phi_S^4+W_{111}\Psi_S^4. \tag{3d}$$

Elastic compliances $s_{ijkl}$, the 4-th order rotostriction coefficients $R_{ijkl}$ and the 6-th order rotostriction coefficients $V_{ijklmn}$ and $W_{ijklmn}$ are written in Vought notations. Note, that 6-th order rotostriction cannot be omitted for correct description of $CaTiO_3$ structural and elastic properties, otherwise it is impossible to describe correctly the structural phase diagram of the bulk material (see **Table C3** in the end of **Appendix C**). Let us underline that the strains (2) are proportional to the product of corresponding rotostriction coefficients, the second and fourth powers of the oxygen displacement components. So, the strains appearance is the typical manifestation of the rotostriction effect.

Note, that the $u^0_{yy}$ and $u^0_{zz}$ are spontaneous strains, which exist in the bulk stress-free single-domain sample. The strain $u^0_{xx}$ contains the spontaneous part and the part proportional to the order parameter variation $\left(\Phi_S^2-\Phi_2^2\right)$, that vanishes far from the twin wall. The appearance of the nonzero out-of-plane strain $u_{zz}(x,0)$ is related with the appearance of a topographic defect ("ditch") on the surface, located at the wall region $x$=0, rather than with limited accuracy of the approximation. Elastic stresses given by Eq.(3) decay by a power law away from wall-surface junction both in x- and z-directions.

Finally, to describe the carrier accumulation at the wall – surface junction, we are also interested in the trace of the strain tensor that can be found as:

$$Tr(u_{ij})\equiv u_{ii}\approx u_{xx}^0+u_{yy}^0+u_{zz}^0-\frac{\delta\sigma\,(1+\nu)(1-2\nu)}{Y}\frac{w\left(2w(w+z)^2+z\left(x^2+(w+z)^2\right)\right)}{\left(x^2+(w+z)^2\right)^2}. \qquad (4)$$

Note, that Eqs.(2)-(4) provide the first-order approximations for strains that do not include polarization-dependent components, e.g. induced electrostriction and flexoelectric coupling. Numerical solution of the nonlinear coupled problem (see **Appendix C** for details) proved that improper polarization induced by the flexo-roto effect [35, 36] in ferroelastic $CaTiO_3$ is relatively small (see section 2.2) and thus the first approximation becomes grounded and consistent. In particular, if we account for the flexoelectric effect contribution in Eqs.(3) it leads to the small second order correction in polarization proportional to the squire of the flexoelectric coupling coefficient.

The strain $u_{xx}(x,z)$, $u_{zz}(x,z)$, $u_{xz}(x,z)$ fields and trace of the strain tensor $\sum_{i=1}^{3}u_{ii}(x,z)$ fields in the vicinity of wall-surface junction for $CaTiO_3$ at room temperature are shown in Fig. 2. Note that the strain $u_{zz}(x,z)$ and the trace $Tr(u_{ij})\equiv\sum_{i=1}^{3}u_{ii}(x,z)$ have a pronounced maximum (~0.5 %) at $x$=0. The strain amplitude decreases and half-width increases away from the surface. The strain profile $u_{xx}(x,z)$ splits in two maxima. As expected, deep in material the strains tend to the spontaneous values $u_{xx}^0$ and $u_{zz}^0$. The strains $u_{xx}(x,z)$ and $u_{zz}(x,z)$ are symmetric with respect to the wall plane $x$=0. The shear strain $u_{xz}(x,z)$ is symmetric with respect to the wall plane $x$=0; it has two maxima which amplitude strongly decreases with z increase. Interestingly, that the scale of the strains amplitude decay is about 10 nm, i.e. well within the applicability limit of mesoscopic theory. The scale is an order of magnitude higher than the domain wall width $w$=0.5 nm, since the strains decrease follows the long-range power law in accordance with Eqs.(2)-(4). Below we will show that the long-range decay could strongly affect on the appearance of the improper polarization and especially on carriers accumulation caused by the wall-surface junction.

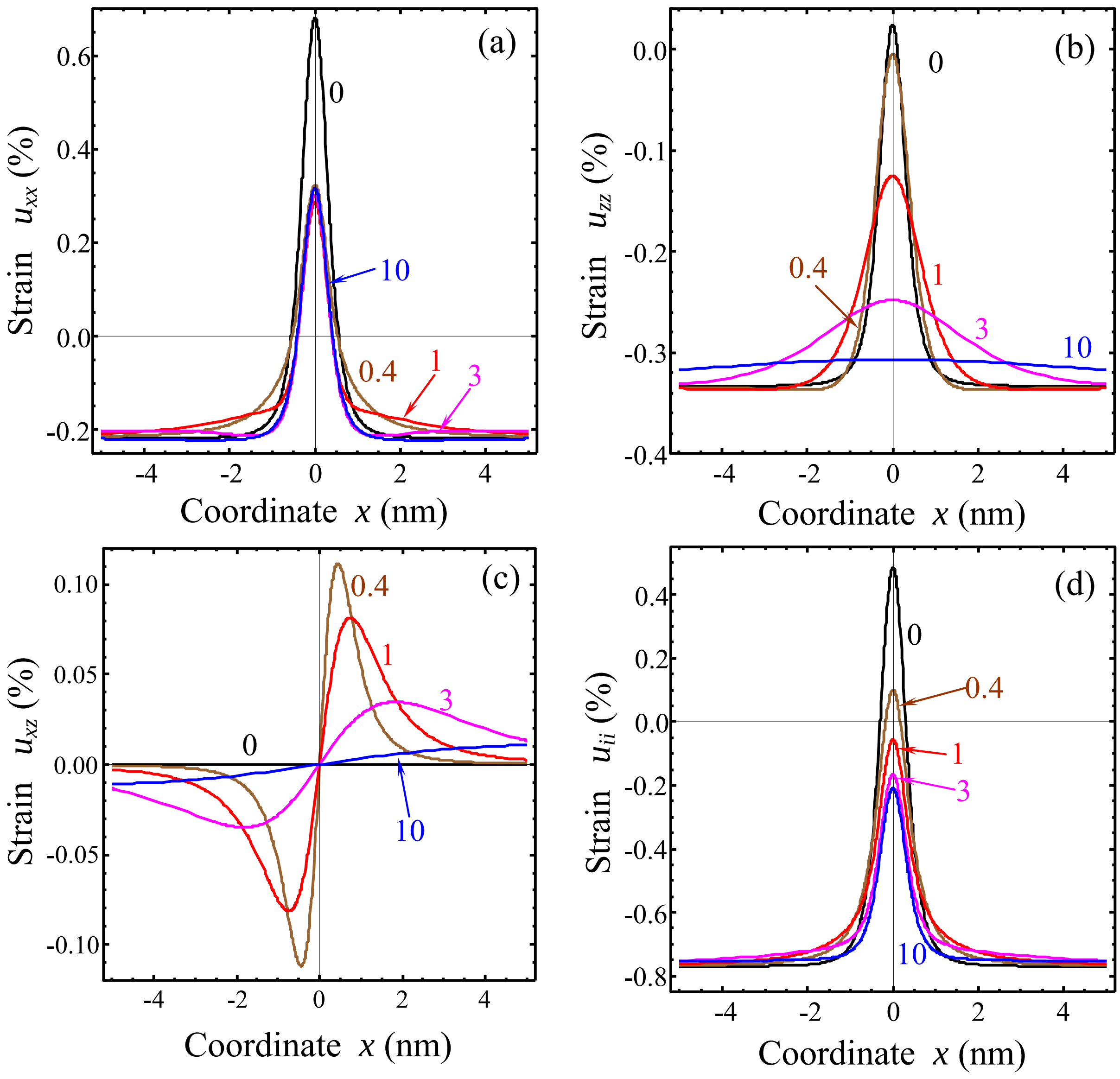

**Figure 2.** (Color online) Strain components $u_{xx}(x,z)$ (a), $u_{zz}(x,z)$ (b), $u_{xz}(x,z)$ (c) and trace of the strain tensor $u_{ii}(x,z)$ (d) vs. the distance $x$ from the twin wall plane $x=0$ calculated at different distances z from the surface (numbers near the curves) for $CaTiO_3$ parameters and room temperature (see **Table C3**). Half- width of the twin wall in the bulk is taken as *w*=0.5 nm.

### 3. Improper ferroelectricity at the twin wall – surface junction

Here, we explore spontaneous polarization induced by flexoelectric coupling in the vicinity of the twin wall-surface junction, as derived in **Appendix C.** Since the values of the gradient coefficients and their anisotropy are yet unknown for $CaTiO_3$, here we did not list the results of numerical simulations based on the free energy minimization. Approximate analytics used hereinafter does not require the knowledge of the coefficients; only the width of domain wall, *w*, is included. However, this is the questionable benefit of the analytical treatment, since the anisotropy of the coefficients essentially influence on the structural order parameter behavior in e.g. $SrTiO_3$ [56, 58]. Flexoelectric coefficients are also unknown for $CaTiO_3$, but the knowledge of their exact values is not very critical for calculations, since the symmetry and their order of magnitude are well-known and relatively high for perovskites.

Polarization fields originating from flexoelectric coupling can be estimated as

$$P_i\left(x,z\right)\approx\alpha_{ij}^{-1}\left(f_{mnjl}\frac{\partial u_{mn}}{\partial x_l}-\frac{\partial\varphi}{\partial x_j}\right). \tag{5}$$

Flexoelectric tensor coefficients are denoted as $f_{mnjl}$, which numerical values $f_{11}$= – 3.24 V, $f_{12}$= 1.44 V, $f_{44}$= 1.08 V were determined experimentally for $SrTiO_3$ [63]; for $BaTiO_3$ $f_{12}$= 450 V was determined experimentally [64, 65] and $f_{11}$=5.12, $f_{12}$=3.32, $f_{44}$=0.045 V were calculated theoretically [66]. Estimations from Kogan model gives $f_{ij}\sim 3.6$V [67].

Electrostatic potential φ can be determined from the Poisson equation, $\varepsilon_0\varepsilon_b\frac{\partial^2\varphi}{\partial x_i^2}=\frac{\partial P_i}{\partial x_i}-e\left(p-n+N_d^+-N_a^-\right)$, where $\varepsilon_0$=8.85×10$^{-12}$ F/m is the universal dielectric constant, $\varepsilon_b$ is the background dielectric permittivity unrelated with the soft mode permittivity $\varepsilon_{ij}^{sm}=\varepsilon_0^{-1}\alpha_{ij}^{-1}$, absolute value of the electron charge $e$=1.6×10$^{-19}$ C; $p(\varphi)$ and $n(\varphi)$ are electrons and holes density; $N_d^+$ and $N_a^-$ are the concentration of ionized acceptors and donors correspondingly. The situation when depolarization effects can be negligibly small in comparison with the flexoelectric polarization $\delta P_i^{flexo}\left(x,z\right)\sim\alpha_{ij}^{-1}f_{mnjl}\frac{\partial u_{mn}}{\partial x_l}\sim\varepsilon_0\varepsilon_{ij}^{sm}f_{mnjl}\frac{\partial u_{mn}}{\partial x_l}$ corresponds to short-circuit electrical boundary conditions. For open-circuit boundary conditions $P_i\left(x,z\right)\sim\frac{\varepsilon_b}{\varepsilon_b+\varepsilon_{ii}^{sm}}f_{mnil}\frac{\partial u_{mn}}{\partial x_l}$, and so it can be much smaller if $\varepsilon_b<<\varepsilon_{ii}^{sm}$. The usage of short-circuit electrical boundary conditions is justified if the surface is covered by the perfectly conductive layer (metallic electrode). Electroded surface is one of typical experimental conditions.

Coefficients $\alpha_{ij}\left(T,x,z\right)$ are affected by elastic fields and biquadratic coupling as [36]:

$$\alpha_{ij}\left(T,x,z\right)=a_{ij}\left(T\right)-q_{ijkl}u_{kl}\left(x,z\right)-\eta_{ijkl}\Phi_k\Phi_l-\xi_{ijkl}\Psi_k\Psi_l \tag{6}$$

Temperature dependence of the coefficient $a_{ii}\left(T\right)$ can be described by the Barrett law for $SrTiO_3$ [68] and $CaTiO_3$ [59]; $q_{ijkl}$ are electrostriction coefficients, $\eta_{ijkl}$ and $\xi_{ijkl}$ are the biquadratic coupling tensor coefficients between the structural and polar order parameter [69, 70]. Elastic stresses induced by the twin wall – surface junction appeared too small to induce ferroelectric polarization in $CaTiO_3$ at room temperature i.e. $\alpha_{ij}\left(T,x,z\right)$ is always positive.

Polarization behavior is primary determined by the strain gradient convolution with the flexoelectric effect tensor in accordance with Eq.(5). Since the strains are proportional to the product of corresponding rotostriction coefficients and the second (or forth) powers of the oxygen displacement components, it can be concluded that in this case the improper polarization Eq. (5) is caused by the flexo-roto effect [35, 36].

The spatial distributions of the in-plane $P_x(x,z)$ and out-of-plane $P_z(x,z)$ polarization components are shown in **Figs.3a,c**. As expected, the $P_x(x,z)$-profiles are antisymmetric and $P_z(x,z)$-profiles are symmetric with respect to the wall plane $x=0$. Polarization profiles have two maxima in the vicinity of the wall plane, which amplitude strongly decreases and half-width increases with z increase more than one lattice constant. The result seems to be in the qualitative agreement with TEM results [50].

Dependencies of the in-plane $P_x(x,z)$ and out-of-plane $P_z(x,z)$ polarization components on the distance z from the $CaTiO_3$ surface are shown in **Figs.3b,d** at different distances x from the twin wall plane $x=0$. Polarization component $P_x(x,z)$ have a pronounced maximum at the surface *z*=0, then it strongly vanishes and diffuses with z increase. Z-dependencies are non-monotonic with a pronounced maximum which amplitude decreases and z-location increases with *x* increase. $P_x(0,z)$ is identically zero as anticipated from the symmetry consideration. $P_z(0,z)$ is maximal under the surface and its value is noticeable (>3 μC/cm$^2$).

Characteristic depth scale of the polarization amplitude decay is about 2 – 3 nm. Polarization value becomes negligibly small at distances ~5 nm from the surface. Similarly to elastic strains, polarization decay obeys the long-range power law in accordance with Eqs.(2)-(4), (5). This behavior is illustrated in log-log plots in **Figs.3e,f** for polarization amplitude $P=\sqrt{P_x^2+P_z^2}$ and tilting angle $\theta=\arccos(P_z/P)$. The spontaneous polarization $P\sim 1$ μC/m$^2$ predicted for the depth of z~100 nm is much higher than the polarization reported for multiferroics-improper ferroelectrics ~1 nC/m$^2$ [71]. The power-law decay of the polarization induced by elastic field at the twin wall-surface junction is fundamentally different from the exponential decay of the polarization induced by the homogeneous de-twinned ferroelastic surface reported earlier [36].

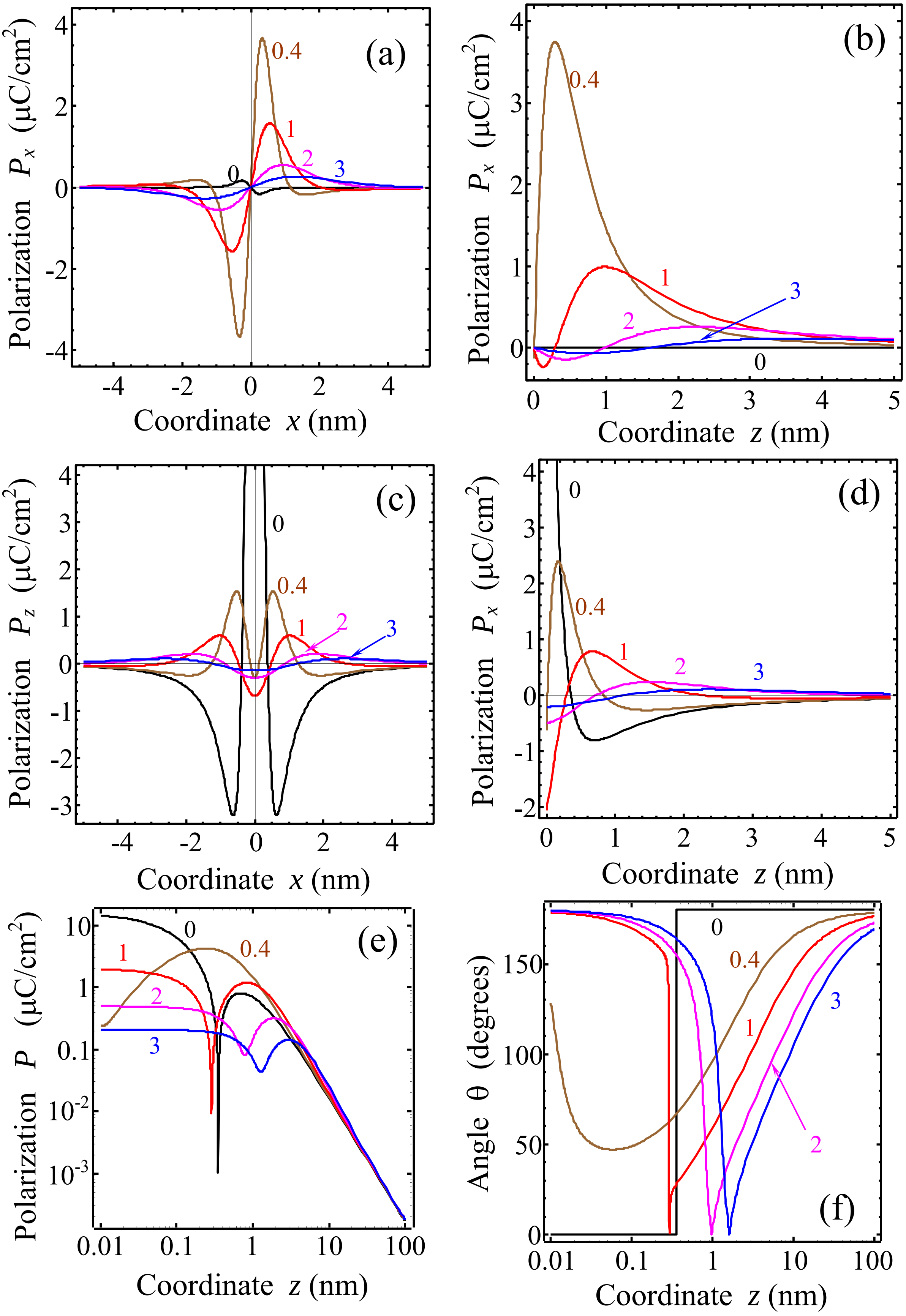

**Figure 3.** (Color online) In-plane $P_x(x,z)$ (a) and out-of-plane $P_z(x,z)$ (c) polarization components vs. the distance $x$ from the twin wall plane $x=0$ calculated at different distances z from the $CaTiO_3$ surface (numbers near the curves). In-plane $P_x(x,z)$ (b) and out-of-plane $P_z(x,z)$ (d) polarization vs.

the distances z from the $CaTiO_3$ surface calculated at different distances $x$ from the twin wall plane $x=0$ (numbers near the curves). Log-log plots of polarization amplitude $P=\sqrt{P_x^2+P_z^2}$ (e) and tilting angle $\theta=\arccos(P_z/P)$ (f). Flexoelectric coefficients $f_{11}$= 16 V, $f_{12}$= -7 eV, $f_{44}$= 5 V. Other parameters are the same as in the **Fig. 2**.

## 4. Carriers accumulation at the twin wall – surface junction

In this section we explore the coupling between order parameter fields and the band structure. In deformation potential theory [72, 73, 74, 75, 76, 77], the strain induced conduction (valence) band edge shift caused by the wall-surface junction is proportional to the strain variation $u_{ij}^S(x,z)=u_{ij}(x,z)-u_{ij}^0(|x|\to\infty)$ given by Eqs.(2), where $u_{ij}^0(|x|\to\infty)$ is the corresponding spontaneous strain given by Eqs.(2) in the limit $\Phi_2^2\to\Phi_S^2$. Thus

$$E_C\left(u_{ij}^S(x,z)\right)=E_{C0}+\Xi_{ij}^C u_{ij}^S(x,z),\qquad E_V\left(u_{ij}^S(x,z)\right)=E_{V0}+\Xi_{ij}^V u_{ij}^S(x,z). \tag{7}$$

where $E_C$ and $E_V$ are the energetic position of the bottom of conduction band and the top of the valence band respectively [78], $\Xi_{ij}^{C,V}$ is a tensor deformation potential of electrons in the conduction ($C$) and valence bands ($V$), where values $E_{C0}=E_C\left(u_{ij}^0(|x|\to\infty)\right)$ and $E_{V0}=E_V\left(u_{ij}^0(|x|\to\infty)\right)$. The symmetry of the deformation potential tensors $\Xi_{ij}^{C,V}$ are rather complex, but it coincide with the crystal spatial symmetry in the Γ-point [75]. Below we use the cubic parent phase approximation of the deformation potential for numerical calculations, i.e. $\Xi_{ij}^{C,V}=-\Xi_d^{C,V}\delta_{ij}$ ($\delta_{ij}$ is a Kroneker-delta symbol), since the spontaneous tetragonal or ortorhombic distortions are absent in the parent phase. Typical absolute values of $\Xi_d^{C,V}$ are ~8 – 12 eV for Ge (see table II in [73]), 21 eV extracted from experimental data for $BiFeO_3$ [79], 8 eV for $SrTiO_3$ estimated from *ab initio* calculations [80], but still poorly studied for both ferroelectrics and ferroelastics. Below we use the value 8 eV [80] for numerical estimations in $CaTiO_3$. Note that both for tetragonal $SrTiO_3$ and orthorhombic $CaTiO_3$ non-diagonal components of deformation potential should be absent.

Electric field $E_i=-\partial\varphi/\partial x_i$ and electrostatic potential φ, are determined self-consistently from the Poisson equation, where the polarization $\delta P_i^{flexo}(x,z)\sim\alpha_{ij}^{-1}f_{mnjl}\frac{\partial u_{mn}}{\partial x_l}$ can be taken as zero order within adopted perturbation approach. Variation of the electric field related with the flexoelectric effect is $\delta E_j^{flexo}\sim-\frac{f_{mnjl}}{1+\varepsilon_0\varepsilon_b\alpha_{11}}\frac{\partial u_{mn}}{\partial x_l}\sim-f_{mnjl}\frac{\partial u_{mn}}{\partial x_l}$. Since $u_{mn}$ is proportional to the proportional to the product of corresponding rotostriction coefficients and the second (or forth) powers of the oxygen displacement components, the field $\delta E_j^{flexo}$ is in fact the so-called flexo-roto field [35, 36]. Corresponding electric potential variation $\delta\varphi_{flexo}$ caused by the field $\delta E_j^{flexo}$ is $\delta\varphi_{flexo}(x,z)=-\int_{-\infty}^{z} dz'\delta E_z(x,z')=f_{mn33}u_{mn}^S(x,z)$.

Allowing for the deformation and electric potential variation and flexoelectric mechanism, local band bending caused by the twin wall-surface junction can be estimated as $\Delta E_n(x,z)=\Xi_d^C u_{ii}^S(x,z)+e\delta\varphi_{flexo}(x,z)$ for electrons and $\Delta E_p(x,z)=-\Xi_d^V u_{ii}^S(x,z)-e\delta\varphi_{flexo}(x,z)$ for holes. The bend bending and changes in electrochemical potentials modulates the densities of free electrons $n(x,z)$ and holes $p(x,z)$ accumulated by the wall – surface junction. The effect can be estimated in the Boltzmann approximation as:

$$n(x,z)\approx n_0 \exp\left(\frac{\Xi_d^C u_{ii}^S(x,z)+ef_{ij33}u_{ij}^S(x,z)}{k_B T}\right), \tag{8a}$$

$$p(x,z)\approx p_0 \exp\left(\frac{-\Xi_d^V u_{ii}^S(x,z)-ef_{ij33}u_{ij}^S(x,z)}{k_B T}\right). \tag{8b}$$

Here $k_B$=1.3807×10$^{-23}$ J/K, $T$ is the absolute temperature, electron charge $e$=1.6×10$^{-19}$ C.

Equations (8a,b) describe the additive effect of deformation potential and flexoelectric coupling into the carriers density and local band bending (see also Eq.(19)-(20) in Refs.[81]). We further note that these equations are valid only as the first order approximation, since polarization and structural order parameter are coupled with each other as well as with elastic fields. Strains also contain polarization-dependent part (via electrostriction effect) and flexoelectric contribution, but all the corrections appeared small and should be considered only in the second order of the perturbation theory.

The static conductivity $\rho(x,z)$ is proportional to the carriers densities (7) as $\rho(x,z)=e\mu_n n(x,z)+e\mu_p p(x,z)$, where $e$ is the electron charge, mobilities $\mu_{n,p}$ are regarded constant. Within the model, the terms $\Xi_d^C u_{ii}^S(x,z)$ and $\Xi_d^V u_{ii}^S(x,z)$ in Eqs.(7) related with deformation potentials explain the ***direct mechanism*** of the twin domain walls static conductivity increase in the semiconducting and insulating ferroelastics and by extension in multiferroics. The trace of the strain tensor $u_{ii}(x,z)$ for considered problem is given by Eq.(4).

As shown previously [35, 36], the flexoelectric coupling leads to the appearance of the inhomogeneous electric fields proportional to the polarization gradient across the wall (flexoelectric field) and to the structural order parameter gradient (roto-flexo field). The fields, which exist at the twin wall – surface junctions, lead to the carrier accumulation via $\delta\varphi_{flexo}(x,z)=f_{mn33}u^{S}_{mn}(x,z)\sim f_{ij33}R_{ijkl}\Phi_k\Phi_l$. Here, we refer to the changes of electrochemical potential due to flexoelectric effect, $\pm ef_{ij33}u^{S}_{ij}(x,z)$ in Eqs.(8), as the ***indirect mechanism*** of domain walls conductivity.

The local bend bending (in linear scale) and carrier density (in log-linear scale) are shown in **Figs.4a** (electrons) and **Fig.4b** (holes). The profiles are symmetric with respect to the wall plane $x=0$ and have a sharp maximum at $x=0$ and small z values. The maximal amplitude decreases and its half-width increases with z increase. The depth profiles of the local bend bending and carrier density are shown in **Figs.4c** (electrons) and **Fig.4d** (holes) at different distances *x* from the twin wall plane $x=0$. Z-dependencies are non-monotonic with a maximum, which amplitude decreases with *x* increase.

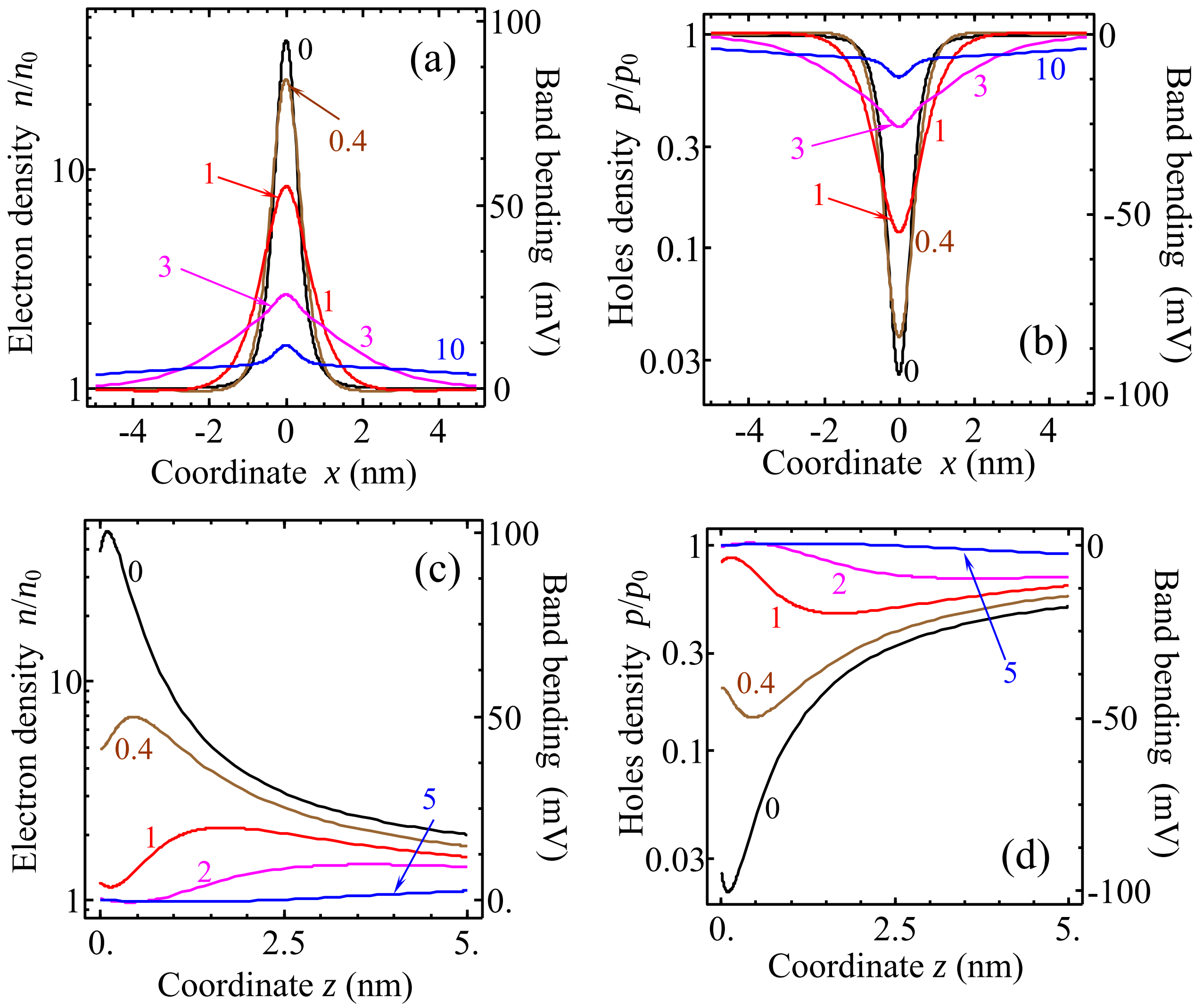

**Figure 4.** (Color online) (a,b) Relative carriers density (left scale) and local band bending (right scale) vs. the distance *x* from the twin wall plane $x=0$ calculated at different distances z from the $CaTiO_3$ surface (numbers near the curves). (c, d) Relative carriers density (left scale) and local band bending

(right scale) vs. the distances z from the $CaTiO_3$ surface calculated at different distances $x$ from the twin wall plane $x=0$ (numbers near the curves). Deformation potentials $\Xi_d^C=\Xi_d^V$=8 eV, flexoelectric coefficients $f_{11}$= 16 V, $f_{12}$= – 7 eV, $f_{44}$= 5 V. Other parameters are the same as in the **Fig. 2**.

For the model case $\Xi_d^C=\Xi_d^V$, the deformation potential mechanism contributes equally to the electron an holes accumulation by the walls (see Eqs.(8)), while the difference between electron and hole band bending originated from the flexoelectric coupling and is equal to $-2ef_{mn33}u_{mn}^S(x,z)$. In numbers, the difference of the accumulated carrier densities appears in the vicinity of the junction. The electrons density near the surface can be about 10 - 40 times higher than the bulk one, while the holes density near the surface can be much smaller than the bulk one for the realistic values of deformation potential $\Xi_d^{C,V}=8$ eV and flexoelectric coefficients $f_{11}$= 16 V, $f_{12}$= – 7 eV, $f_{44}$= 5 V. Hence, the flexoelectric coupling sign is primary responsible for the twin walls n-type (or p-type) conductivity in ferroelastics proper-semiconductors. Consequently the strength of the indirect contribution can be estimated from the difference of the electrons and holes densities accumulate by the twin walls. For material parameters used, the relative contribution of deformation potential and flexoelectric coupling into the carrier accumulation are of the same order. For materials with weak flexoelectric coupling (such as $SrTiO_3$ with $f_{11}$= 1.6 V, $f_{12}$= – 0.7 eV, $f_{44}$= 0.5 V [63]) deformation potential contribution should dominate. It is seen from the contour maps of the electrons (**Fig.5a**) and holes (**Fig.5b**) densities that characteristic scale of the carrier densities variation is about 5 nm.

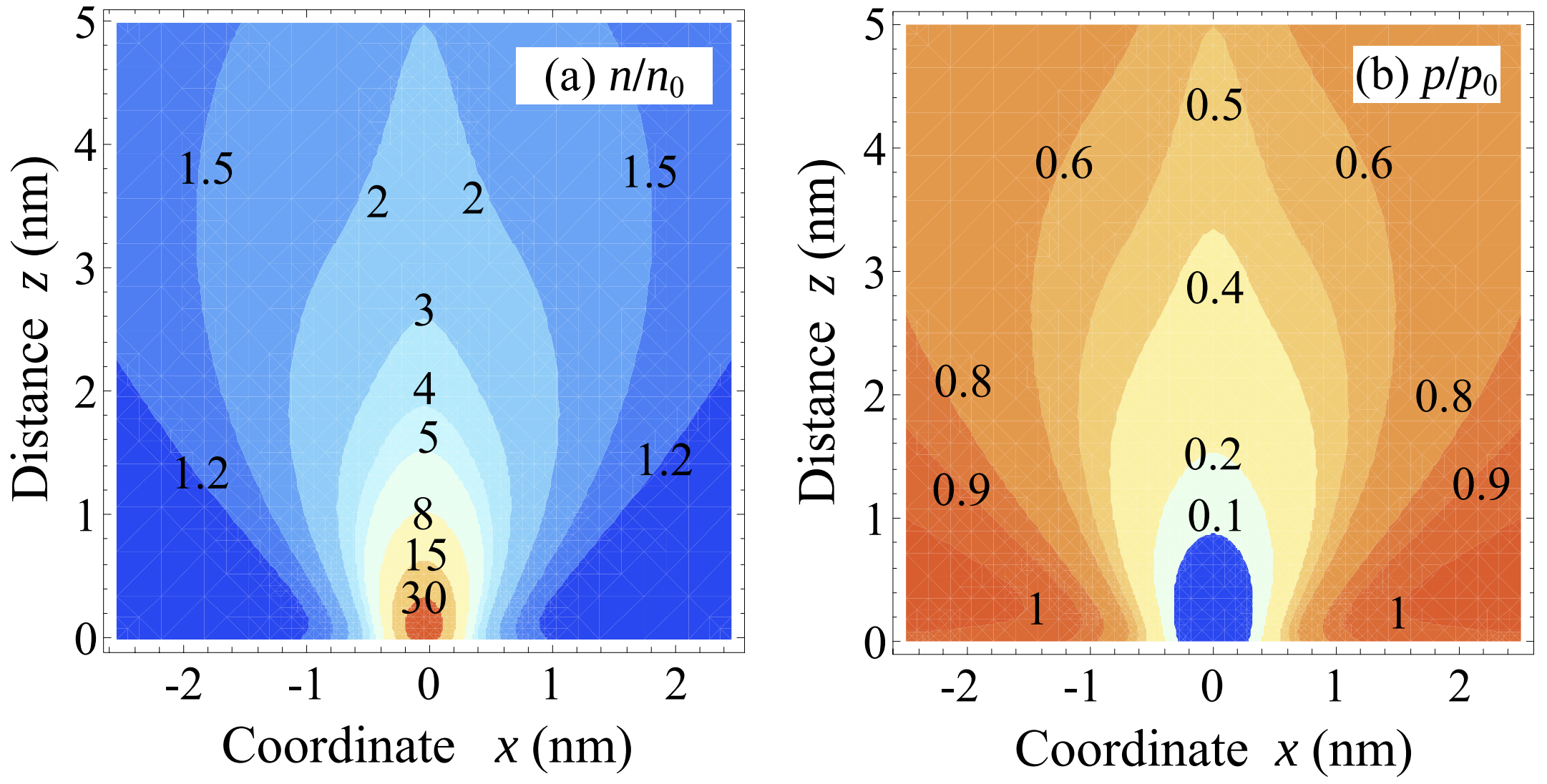

**Figure 5.** (Color online) Contour maps of the electronic (a) and holes (b) density in the vicinity of the twin wall – surface junction in $CaTiO_3$. The twin wall plane is $x=0$. Other parameters are the same as in the **Fig. 4**.

### 5. Vacancies segregation at the twin wall – surface junction

Similarly to the electrons accumulation and holes depletion, the twin wall – surface junction in $CaTiO_3$ (or $SrTiO_3$) can accumulate donors (oxygen vacancies) or acceptors (titanium vacancies). Here, the Vegard expansion (elastic dipole) tensor $\beta_{jk}^{a,d}$ plays the same role in the vacancies segregation as the deformation potential tensor in the electron accumulation. The structure of Vegard expansion tensor is controlled by the symmetry (crystalline or Curie group symmetry) of the material. For isotropic or cubic media it is diagonal and reduces to scalar: $\beta_{ij}^{a,d} = \beta_{ii}^{a,d}\delta_{ij}$ [82].

Using analytical expressions for single-ionized donors concentration $N_d^+ \approx N_{d0}^+ \exp\left(\left(\beta_{jk}^d u_{jk} - e\varphi\right)/k_B T\right)$ and $N_a^- \approx N_{a0}^- \exp\left(\left(\beta_{jk}^a u_{jk} + e\varphi\right)/k_B T\right)$ for single-ionized acceptors concentration (derived in Ref. [81]) and the expression for potential variation $\delta\varphi_{flexo}(x,z) = f_{mn33} u_{mn}^S(x,z)$, corresponding analytical expressions have the form:

$$N_d^+(x,z) \approx N_{d0}^+ \exp\left(\frac{\beta_{ii}^d u_{ii}^S(x,z) - e f_{ij33} u_{ij}^S(x,z)}{k_B T}\right), \tag{9a}$$

$$N_a^-(x,z) \approx N_{a0}^- \exp\left(\frac{\beta_{ii}^a u_{ii}^S(x,z) + e f_{ij33} u_{ij}^S(x,z)}{k_B T}\right). \tag{9b}$$

Here $N_{d0}^+$ and $N_{a0}^-$ are the concentration of single-ionized vacancies in the bulk. For neutral vacancies corresponding expressions are $N_d^0(x,z) \approx N_{d0} \exp\left(\frac{\beta_{ii}^d u_{ii}^S(x,z)}{k_B T}\right)$ and $N_a^0(x,z) \approx N_{a0} \exp\left(\frac{\beta_{ii}^a u_{ii}^S(x,z)}{k_B T}\right)$.

For numerical estimations of the O vacancies accumulation we use the anisotropic values of the elastic dipole tensor $\beta_{22}^d = \beta_{33}^d = -2.13$ eV, $\beta_{11}^d = 4.53$ eV (vacancy orientation $V_{OX}^-$), $\beta_{11}^d = \beta_{33}^d = -2.13$ eV, $\beta_{22}^d = 4.53$ eV (vacancy orientation $V_{OY}^-$), $\beta_{11}^d = \beta_{22}^d = -2.13$ eV, $\beta_{33}^d = 4.53$ eV (vacancy orientation $V_{OZ}^-$) and the isotropic average $\beta_{ii}^d = 0.09$ eV ($V_O^-$ vacancy) as calculated by Freedman et al [82]. For Ti vacancy we use the values $\beta_{11}^a = \beta_{22}^a = \beta_{33}^a = 28$ eV [82]. Profiles of the electrochemical potential $\chi_d(x,z) = \beta_{ii}^d u_{ii}^S(x,z) - e f_{ij33} u_{ij}^S(x,z)$ and concentration of O-vacancies (in log-linear scale) are shown in **Figs. 6** at different distances $z$ from the $CaTiO_3$ surface and room temperature. Vacancies concentration can be higher than the bulk one in 2-7 times (compare the result with [45]). In the surface plane, the profiles are symmetric with respect to the wall plane $x = 0$ and have a sharp maximum (or minimum) at $x = 0$. The maximum (minimum) amplitude decreases and its half-width increases with z increase. For z ≥ 3 nm the vacancies behavior are the same as around the twin wall in the bulk.

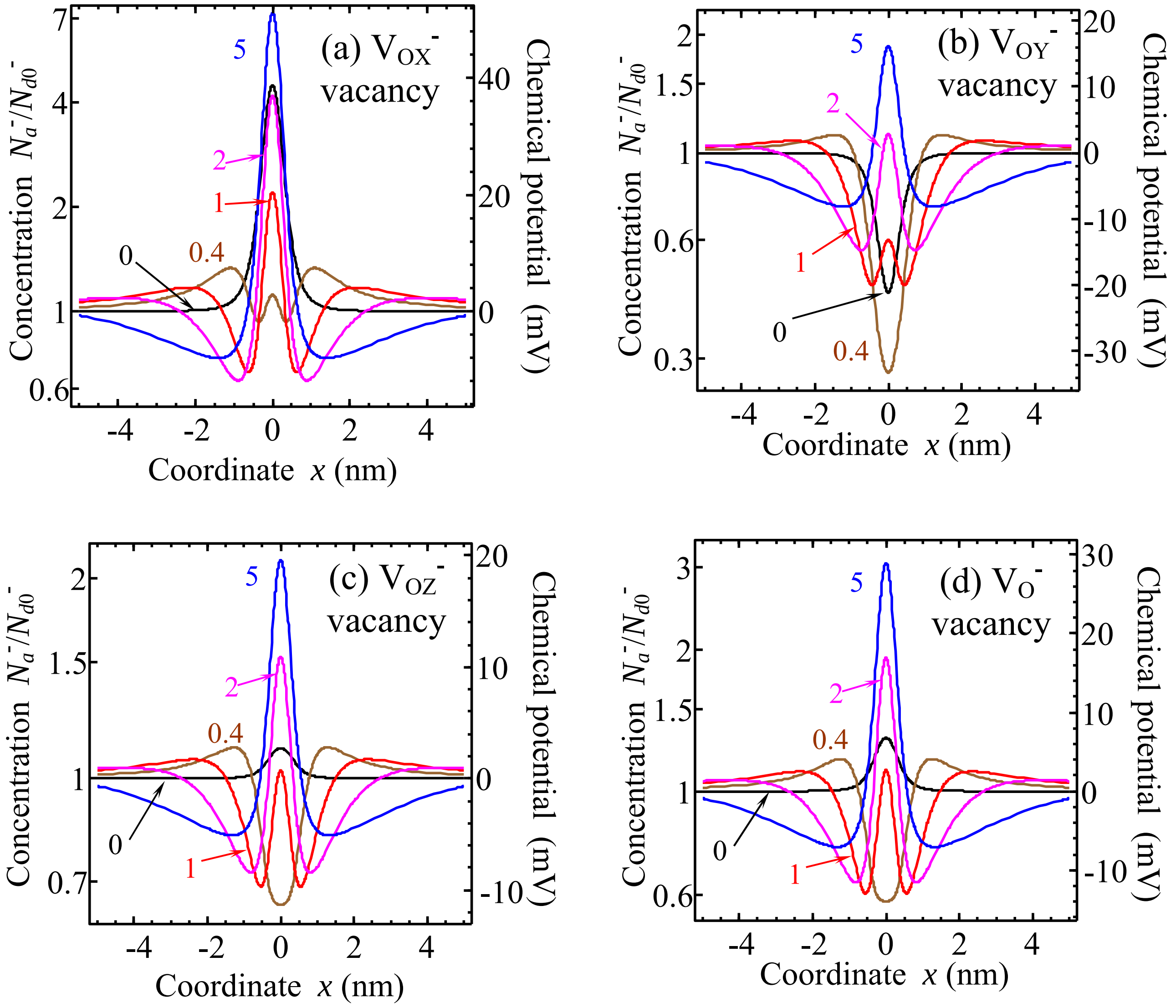

**Figure 6.** (Color online) (a,b) Relative concentration of single-ionized O vacancies (left scale) and their electrochemical potential variation (right scale) vs. the distance $x$ from the twin wall plane $x = 0$ calculated at different distances z from the $CaTiO_3$ surface (numbers near the curves). Elastic dipole tensor $\beta_{22}^d = \beta_{33}^d = -2.13$ eV, $\beta_{11}^d = 4.53$ eV for vacancy orientation $V_{OX}^-$ (a); $\beta_{11}^d = \beta_{33}^d = -2.13$ eV, $\beta_{22}^d = 4.53$ eV for vacancy orientation $V_{OY}^-$ (b); $\beta_{11}^d = \beta_{22}^d = -2.13$ eV, $\beta_{33}^d = 4.53$ eV for vacancy orientation $V_{OZ}^-$ (c); $\beta_{ii}^d = 0.09$ eV for $V_O^-$ vacancy (d). Other parameters are the same as in the **Fig. 4**.

Spatial distribution of $V_{OX}^-$ vacancies concentration in the vicinity of the twin wall – surface junction in $CaTiO_3$ is shown in **Fig. 7.** It is seen from the figure that the distribution of $V_{OX}^-$ vacancies is rather complex and strongly affected by the surface.

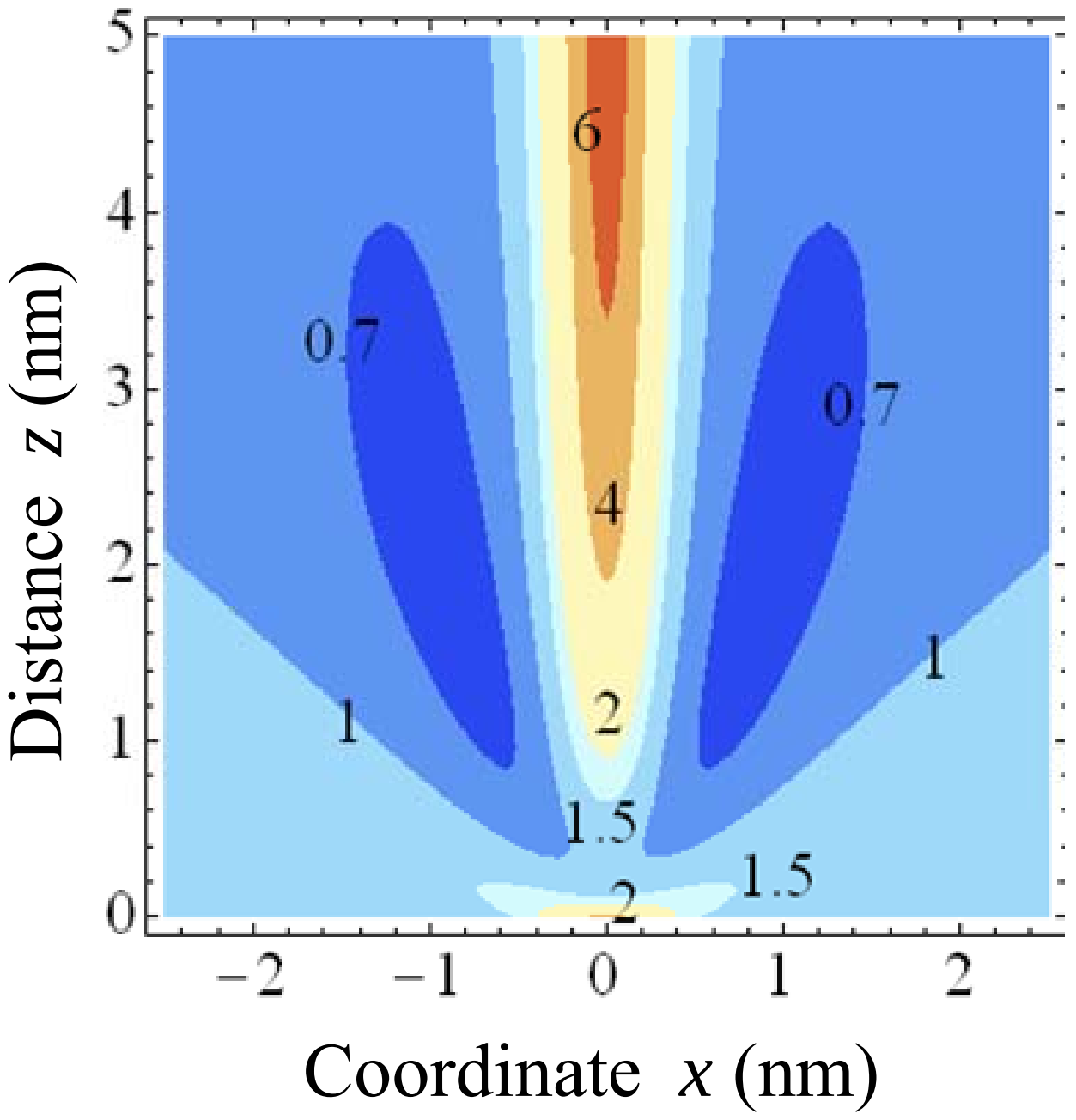

**Figure 7.** (Color online) Contour maps of the $V_{OX}^{-}$ vacancies concentration in the vicinity of the twin wall – surface junction in $CaTiO_3$. The twin wall plane is $x = 0$. Other parameters are the same as in the **Fig. 6**.

For used material parameters the relative contribution of Vegard expansion and flexoelectric coupling into the charged vacancies segregation are of the same order. If the charged vacancies are mobile, they can also contribute the wall conductivity.

Note, that single-charged Ti vacancies concentration near the domain wall can be about $10^3$ times higher than the bulk one. This is related with high values of elastic dipole for theses defects [82]. Since the equilibrium “bulk” concentration of Ti vacancies should be much smaller than for the oxygen vacancies we showed the figures for the oxygen vacancies only. Neural vacancies and bi-vacancies cannot change the wall conductivity, but they can be the most thermodynamically stable defects.

## 6. Summary

Inhomogeneous elastic strains, which exist due to the rotostriction in the vicinity of the twin walls – surface junctions in ferroelastics, can strongly affect their electronic properties. In particular, the strains change the band structure at the wall – surface junction via the deformation potential, rotostriction and flexoelectric coupling mechanisms. The calculated decrease of the local band gap can be considered as the *direct mechanism* of the uncharged domain walls conductivity increase in the ferroelastics ($CaTiO_3$, $EuTiO_3$, $SrTiO_3$, etc), by extension in other multiferroics ($BiFeO_3$) and semiconducting ferroelectrics ($PbTiO_3$, $BaTiO_3$, etc). Flexoelectric and rotostriction couplings lead to the appearance of the inhomogeneous electric field, named roto-flexo field [35, 36], proportional to the structural order parameter gradient across the wall. The fields, which are localized at the domain wall plane and lead to the carrier accumulation in the wall region, are considered as *indirect mechanism* of the uncharged domain walls conductivity (see **Fig. 8**). Comparison of the direct and indirect mechanisms contribution to the twin wall static conductivity demonstrated their complex interplay in $CaTiO_3$. The conductivity response of the twin wall – surface junction predicted here can be verified by SPM, since the junction is the inherent part of conduction path involved in the probing.

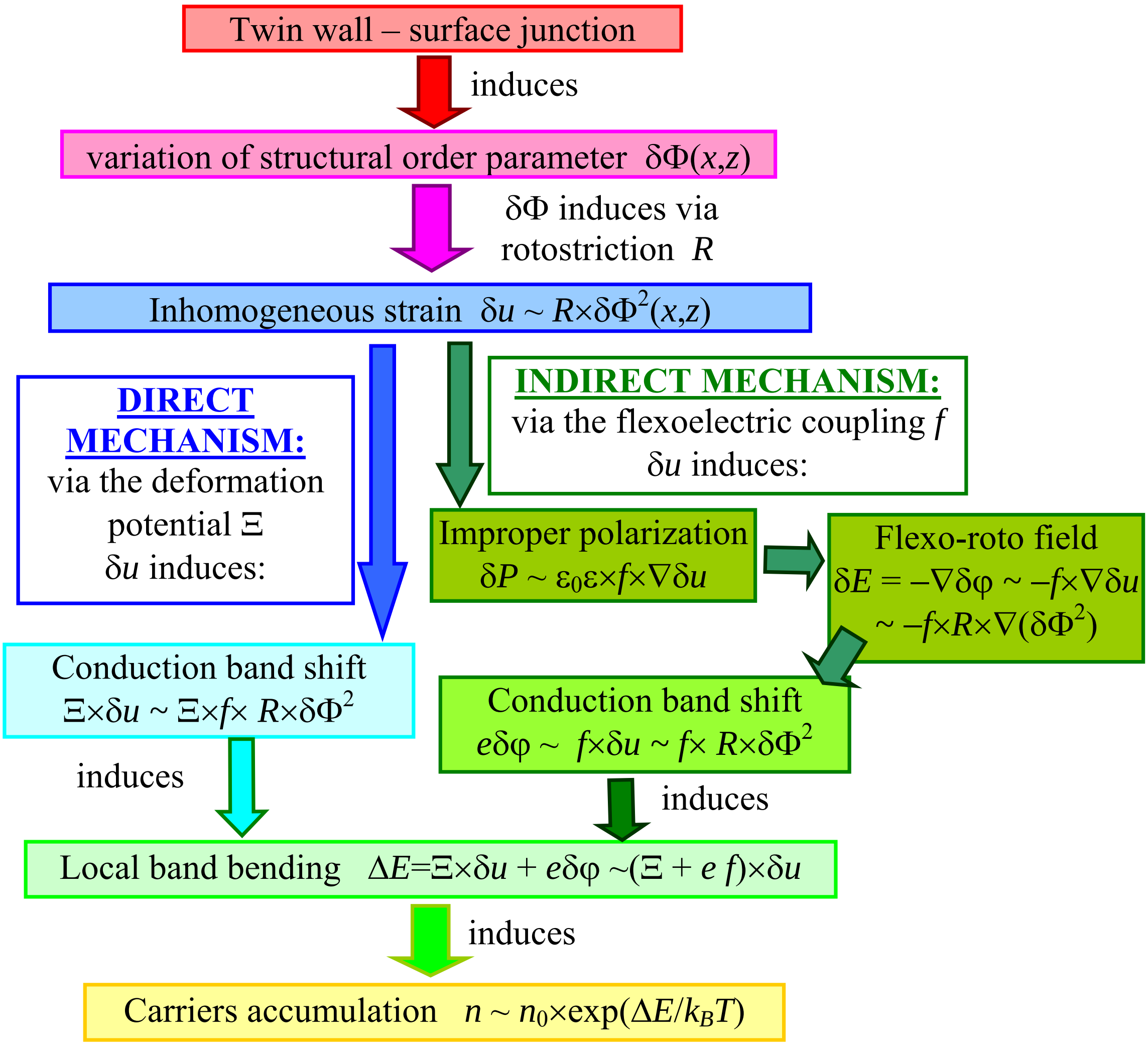

**Figure 8.** (Color online) Origin of the twin wall – surface junction spontaneous polarization and electro-conductivity: direct and indirect mechanisms.

**Acknowledgements**

E.A.E. and A.N.M. gratefully acknowledge multiple discussions with N.V. Morozovskii (NAS Ukraine), Nicole Benedek (Cornell University) and Karin Rabe (Rutgers University). V.G. and L.Q.C. acknowledges NSF-DMR- 0908718 and DMR-0820404 funds. E.A.E. and A.N.M. are thankful to NAS Ukraine and NSF-DMR-0908718 for support. Research supported (SVK and AB) by the U.S. Department of Energy, Basic Energy Sciences, Materials Sciences and Engineering Division.

## Supplemental Materials

### Appendix A

Green's function tensor for semi-infinite isotropic elastic half-plane is given by Lur'e [83] and Landau and Lifshitz [84] as:

$$G_{ij}(x_1,x_2,\xi_3)=\begin{cases}\dfrac{1+\nu}{2\pi Y}\left[\dfrac{\delta_{ij}}{R}+\dfrac{(x_i-\xi_i)(x_j-\xi_j)}{R^3}+\dfrac{1-2\nu}{R+\xi_3}\left(\delta_{ij}-\dfrac{(x_i-\xi_i)(x_j-\xi_j)}{R(R+\xi_3)}\right)\right] & i,j\neq 3\\ \dfrac{(1+\nu)(x_i-\xi_i)}{2\pi Y}\left(\dfrac{-\xi_3}{R^3}-\dfrac{(1-2\nu)}{R(R+\xi_3)}\right) & i=1,2 \quad and \quad j=3\\ \dfrac{(1+\nu)(x_j-\xi_j)}{2\pi Y}\left(\dfrac{-\xi_3}{R^3}+\dfrac{(1-2\nu)}{R(R+\xi_3)}\right) & j=1,2 \quad and \quad i=3\\ \dfrac{1+\nu}{2\pi Y}\left(\dfrac{2(1-\nu)}{R}+\dfrac{\xi_3{}^2}{R^3}\right) & i=j=3\end{cases} \tag{A.1}$$

Here $R=\sqrt{(x_1-\xi_1)^2+(x_2-\xi_2)^2+\xi_3^2}$ is radius vector, $Y$ is Young's modulus, and $\nu$ is the Poisson ratio. Stiffness tensor $c_{kjmn}$ corresponds to the elastically isotropic medium

$$c_{klmn}=\frac{Y}{2(1+\nu)}\left[\frac{2\nu}{1-2\nu}\delta_{kl}\delta_{mn}+\delta_{km}\delta_{ln}+\delta_{kn}\delta_{lm}\right].$$

### Appendix B

***Case I. 90-degree twins in ferroelastics with tetragonal symmetry***

Introducing 45°-rotated coordinate system [85] with one axis, perpendicular to twin wall, $z\equiv x_3$, and two other rotated axes, $x=(x_1+x_2)/\sqrt{2}$ , $y=(-x_1+x_2)/\sqrt{2}$ , distributions of elastic stresses unperturbed by the surface influence have the form [86]:

$$\sigma_{22}^0=\frac{s_{11}U_2-s_{12}U_3}{s_{11}\tilde{s}_{11}-s_{12}^2}, \qquad \sigma_{33}^0=\frac{\tilde{s}_{11}U_3-s_{12}U_2}{s_{11}\tilde{s}_{11}-s_{12}^2}, \qquad \sigma_{11}^0=\sigma_{13}^0=\sigma_{12}^0=\sigma_{32}^0=0. \tag{B.1}$$

Using the decoupling approximation on polarization (i.e. neglecting electrostriction and flexoelectric effect), functions $U_{2,3}$ determined by the structural order parameter (oxygen displacements $\Phi_1$ and $\Phi_2$, see **Fig.1b**) distributions have the form:

$$U_2\approx\tilde{R}_{11}\left(\frac{\Phi_S^2}{2}-\tilde{\Phi}_2^2\right)+\tilde{R}_{12}\left(\frac{\Phi_S^2}{2}-\tilde{\Phi}_1^2\right), \tag{B.2}$$

$$U_3\approx R_{12}\left(\Phi_S^2-\tilde{\Phi}_2^2-\tilde{\Phi}_1^2\right). \tag{B.3}$$

Here we used the tensor components in the new reference frame for elastic compliances $\tilde{s}_{11}=(s_{11}+s_{12}+s_{44}/2)/2$ , $\tilde{s}_{12}=(s_{11}+s_{12}-s_{44}/2)/2$ , rotostriction $\tilde{R}_{11}=(R_{11}+R_{12}+R_{44}/2)/2$ , $\tilde{R}_{12}=(R_{11}+R_{12}-R_{44}/2)/2$. In order to obtain analytical results (Pade approximations) structural order parameters were used in the form unperturbed by the surface, namely $\tilde{\Phi}_1=\left(\Phi_S/\sqrt{2}\right)\tanh(\tilde{x}_1/w)$ and $\tilde{\Phi}_2=\left(\Phi_S/\sqrt{2}\right)\left(1+a\cosh^{-2}(\tilde{x}_1/w)\right)\approx\Phi_S/\sqrt{2}$ (the amplitude $a$ is typically much smaller than unity), where the structural wall width $w$ is about several lattice constant.

***Case II. 90-degree twins in ferroelastics with orthorhombic symmetry***

For the considered case of nonzero structural order parameter components $\Phi_1$, $\Phi_2$ and $\Psi_3$ (see **Fig.1b**) inhomogeneous stress can be written in the form:

$$\sigma_{22}^0=\frac{s_{11}U_2-s_{12}U_3}{s_{11}^2-s_{12}^2},\ \sigma_{33}^0=\frac{s_{11}U_3-s_{12}U_2}{s_{11}^2-s_{12}^2},\ \sigma_{11}^0=\sigma_{13}^0=\sigma_{12}^0=\sigma_{32}^0=0. \tag{B.4}$$

Functions $U_{2,3}$ are determined by the structural order parameter components as:

$$\begin{aligned}U_2=&R_{11}\left(\Phi_S^2-\Phi_2^2\right)+R_{12}\left(\Phi_S^2-\Phi_1^2\right)+Z_{12}\left(\Psi_S^2-\Psi_3^2\right)+\\&+V_{111}\left(\Phi_S^4-\Phi_2^4\right)+6V_{112}\left(\Phi_S^4-\Phi_1^2\Phi_2^2\right)+V_{122}\left(\Phi_S^4-\Phi_1^4\right)+W_{122}\left(\Psi_S^4-\Psi_3^4\right)\end{aligned}, \tag{B.5a}$$

$$\begin{aligned}U_3=&R_{12}\left(2\Phi_S^2-\left(\Phi_1^2+\Phi_2^2\right)\right)+Z_{11}\left(\Psi_S^2-\Psi_3^2\right)+\\&+V_{122}\left(2\Phi_S^4-\Phi_1^4-\Phi_2^4\right)+6V_{123}\left(\Phi_S^4-\Phi_1^2\Phi_2^2\right)+W_{111}\left(\Psi_S^4-\Psi_3^4\right)\end{aligned}, \tag{B.5b}$$

structural order parameters were used in the form unperturbed by the surface, $\Phi_1\approx\Phi_S\left(1+a\cosh^{-2}(x_1/w)\right)$, $\Phi_2\approx\Phi_S\tanh(x_1/w)$, $\Psi_3\approx\Psi_S\left(1+b\cosh^{-2}(x_1/w)\right)$. The amplitudes $a$ and $b$ appeared much smaller than unity. For the specific case DW should be perpendicular to $[100]$ or $[010]$ directions.

The "true" stresses $\sigma_{kl}(x,z)$ can be calculated from Eq.(1) and Eq.(B.4) using the perturbation approach as $\sigma_{kl}(x,z)=\sigma_{kl}^0+\frac{c_{klij}}{2}\left(\frac{\partial u_i^S}{\partial x_j}+\frac{\partial u_j^S}{\partial x_i}\right)$. After lengthy calculations we obtained Pade approximations for nonzero stresses:

$$\sigma_{11}(x,z)\approx-\frac{\delta\sigma\, w\left(\left(w^2+x^2\right)(w+z)^3+x^2z\left(x^2+3w(w+z)\right)\right)}{\left(x^2+(w+z)^2\right)^3}, \tag{B.6a}$$

$$\sigma_{22}(x,z)\approx\sigma_{22}^0-\nu\frac{\delta\sigma\, w\left(2w(w+z)^2+z\left(x^2+(w+z)^2\right)\right)}{\left(x^2+(w+z)^2\right)^2}, \tag{B.6b}$$

$$\sigma_{13}(x,z)\approx-\frac{\delta\sigma\, wxz\left(4w(w+z)^2+z\left(x^2+(w+z)^2\right)\right)}{\left(x^2+(w+z)^2\right)^3}, \tag{B.6c}$$

$$\sigma_{33}(x,z)\approx\sigma_{33}^0-\frac{\delta\sigma\, w\left(4wz(w+z)^3+\left(x^2+(w+z)^2\right)\left(w^3+z^3\right)\right)}{\left(x^2+(w+z)^2\right)^3}. \tag{B.6d}$$

Poisson ratio is ν, $w$ is the intrinsic width of the twin wall in the bulk, coordinates $x_1=x$, $x_2=y$ and $x_3=z$. The amplitude $\delta\sigma$ is different for the twins in tetragonal and orthorhombic ferroelastics:

$$\delta\sigma=\frac{\tilde{s}_{11}R_{12}-s_{12}\tilde{R}_{12}}{s_{11}\tilde{s}_{11}-s_{12}^2}\frac{\Phi_S^2}{2}, \qquad \text{(tetragonal)} \tag{B.7a}$$

$$\delta\sigma\approx\frac{s_{11}\left(R_{12}+(V_{122}+6V_{112})\Phi_S^2\right)-s_{12}\left(R_{11}+(V_{111}+6V_{112})\Phi_S^2\right)}{s_{11}^2-s_{12}^2}\Phi_S^2. \qquad \text{(orthorhombic)} \tag{B.7b}$$

In Eqs.(B.7) we used the tensor components in the new reference frame for elastic compliances $\tilde{s}_{11}=(s_{11}+s_{12}+s_{44}/2)/2$, $\tilde{s}_{12}=(s_{11}+s_{12}-s_{44}/2)/2$, rotostriction $\tilde{R}_{11}=(R_{11}+R_{12}+R_{44}/2)/2$, $\tilde{R}_{12}=(R_{11}+R_{12}-R_{44}/2)/2$.

Note that $\sigma_{33}(x,0)$ is not identically zero. The deviation is related with the appearance of a ditch on the surface, located at the wall region $x$=0, rather than with limited accuracy of the approximation. Elastic stresses given by Eq.(B.6) decay in accordance with power law with both $z$ and $x$ increase.

For the carrier accumulation by the wall – surface junction trace of stress tensor is important, that has the form:

$$\sigma_{11}(x,z)+\sigma_{22}(x,z)+\sigma_{33}(x,z)\approx\sigma_{22}^0+\sigma_{33}^0-(1+\nu)\frac{\delta\sigma\, w\left(2w(w+z)^2+z\left(x^2+(w+z)^2\right)\right)}{\left(x^2+(w+z)^2\right)^2} \tag{B.8}$$

Where $\sigma_{22}^0+\sigma_{33}^0=\dfrac{U_3+U_2}{s_{11}+s_{11}}$ (see Eqs.(B.4)-(B.5)).

## Appendix C. Free energy functional and material parameters of $CaTiO_3$

### C.1. Free energy functional and related equations

Polarization dependent free energy density:

$$\begin{aligned}G_P&=\alpha_1(T)\left(P_1^2+P_2^2+P_3^2\right)+\alpha_{11}\left(P_1^2+P_2^2+P_3^2\right)^2+\alpha_{12}\left(P_1^4+P_2^4+P_3^4\right)+\\&+\alpha_{11}\left(P_1^2+P_2^2+P_3^2\right)^3+\alpha_{112}\left(P_1^2+P_2^2+P_3^2\right)\left(P_1^4+P_2^4+P_3^4\right)+\alpha_{122}P_1^2P_2^2P_3^2-P_iE_i\end{aligned} \tag{C.1}$$

SOP dependent free energy density:

$$\begin{aligned}G_{SOP}&=\beta_1(T)\left(\Phi_1^2+\Phi_2^2+\Phi_3^2\right)+\beta_{11}\left(\Phi_1^2+\Phi_2^2+\Phi_3^2\right)^2+\beta_{12}\left(\Phi_1^4+\Phi_2^4+\Phi_3^4\right)+\\&+\beta_{111}\left(\Phi_1^2+\Phi_2^2+\Phi_3^2\right)^3+\beta_{112}\left(\Phi_1^2+\Phi_2^2+\Phi_3^2\right)\left(\Phi_1^4+\Phi_2^4+\Phi_3^4\right)+\beta_{122}\Phi_1^2\Phi_2^2\Phi_3^2+\\&+\gamma_1(T)\left(\Psi_1^2+\Psi_2^2+\Psi_3^2\right)+\gamma_{11}\left(\Psi_1^2+\Psi_2^2+\Psi_3^2\right)^2+\gamma_{12}\left(\Psi_1^4+\Psi_2^4+\Psi_3^4\right)+\\&+\gamma_{111}\left(\Psi_1^2+\Psi_2^2+\Psi_3^2\right)^3+\gamma_{112}\left(\Psi_1^2+\Psi_2^2+\Psi_3^2\right)\left(\Psi_1^4+\Psi_2^4+\Psi_3^4\right)+\gamma_{122}\Psi_1^2\Psi_2^2\Psi_3^2\end{aligned} \tag{C.2}$$

Coefficients $\alpha_1(T)$, $\beta_1(T)$ and $\gamma_1(T)$ can be fitted by the Barrett law

$$a_i(T)=\alpha_T T_q^{(E)}\left(\coth\left(T_q^{(E)}/T\right)-\coth\left(T_q^{(E)}/T_0^{(E)}\right)\right), \tag{C.3a}$$

$$\beta_1(T)=\beta_T\Theta_{s2}\left(\coth\left(\Theta_{s2}/T\right)-\coth\left(\Theta_{s2}/T_2\right)\right), \tag{C.3b}$$

$$\gamma_1(T)=\gamma_T\Theta_{s3}\left(\coth\left(\Theta_{s3}/T\right)-\coth\left(\Theta_{s3}/T_3\right)\right). \tag{C.3c}$$

We consider biquadratic coupling [87, 88] between SOP and polarization is

$$\begin{aligned}
G_{SOPP} &= -t_{11}\left(\Phi_1^2P_1^2+\Phi_2^2P_2^2+\Phi_3^2P_3^2\right)-t_{12}\left(\Phi_3^2\left(P_1^2+P_2^2\right)+\Phi_1^2\left(P_2^2+P_3^2\right)+\Phi_2^2\left(P_3^2+P_1^2\right)\right)-\\
&-t_{44}\left(\Phi_2\Phi_3P_2P_3+\Phi_3\Phi_1P_3P_1+\Phi_1\Phi_2P_1P_2\right)-\\
&-\kappa_{11}\left(\Psi_1^2P_1^2+\Psi_2^2P_2^2+\Psi_3^2P_3^2\right)-\kappa_{12}\left(\Psi_3^2\left(P_1^2+P_2^2\right)+\Psi_1^2\left(P_2^2+P_3^2\right)+\Psi_2^2\left(P_3^2+P_1^2\right)\right)-\\
&-\kappa_{44}\left(\Psi_2\Psi_3P_2P_3+\Psi_3\Psi_1P_3P_1+\Psi_1\Psi_2P_1P_2\right)-\\
&-\mu_{11}\left(\Phi_1^2\Psi_1^2+\Phi_2^2\Psi_2^2+\Phi_3^2\Psi_3^2\right)-\mu_{12}\left(\Psi_3^2\left(\Phi_1^2+\Phi_2^2\right)+\Psi_1^2\left(\Phi_2^2+\Phi_3^2\right)+\Psi_2^2\left(\Phi_3^2+\Phi_1^2\right)\right)-\\
&-\mu_{44}\left(\Psi_2\Psi_3\Phi_2\Phi_3+\Psi_3\Psi_1\Phi_3\Phi_1+\Psi_1\Psi_2\Phi_1\Phi_2\right)
\end{aligned} \tag{C.4}$$

Electrostriction and rotostriction coupling energy (2nd order striction)

$$\begin{aligned}
G_2^{striction} &= -Q_{11}\left(\sigma_1P_1^2+\sigma_2P_2^2+\sigma_3P_3^2\right)-Q_{12}\left[\sigma_1\left(P_2^2+P_3^2\right)+\sigma_2\left(P_3^2+P_1^2\right)+\sigma_3\left(P_1^2+P_2^2\right)\right]\\
&-Q_{44}\left(\sigma_4P_2P_3+\sigma_5P_3P_1+\sigma_6P_1P_2\right)-\\
&-\frac{1}{2}s_{11}\left(\sigma_1^2+\sigma_2^2+\sigma_3^2\right)-s_{12}\left(\sigma_1\sigma_2+\sigma_2\sigma_3+\sigma_3v_1\right)-\frac{1}{2}s_{44}\left(\sigma_4^2+\sigma_5^2+\sigma_6^2\right)-\\
&-R_{11}\left(\sigma_1\Phi_1^2+\sigma_2\Phi_2^2+\sigma_3\Phi_3^2\right)-R_{12}\left[\sigma_1\left(\Phi_2^2+\Phi_3^2\right)+\sigma_2\left(\Phi_3^2+\Phi_1^2\right)+\sigma_3\left(\Phi_1^2+\Phi_2^2\right)\right]\\
&-R_{44}\left(\sigma_4\Phi_2\Phi_3+\sigma_5\Phi_3\Phi_1+\sigma_6\Phi_1\Phi_2\right)-\\
&-Z_{11}\left(\sigma_1\Psi_1^2+\sigma_2\Psi_2^2+\sigma_3\Psi_3^2\right)-Z_{12}\left[\sigma_1\left(\Psi_2^2+\Psi_3^2\right)+\sigma_2\left(\Psi_1^2+\Psi_3^2\right)+\sigma_3\left(\Psi_1^2+\Psi_2^2\right)\right]\\
&-Z_{44}\left(\sigma_4\Psi_2\Psi_3+\sigma_5\Psi_1\Psi_3+\sigma_6\Psi_1\Psi_2\right)
\end{aligned} \tag{C.5a}$$

The forth-rank electrostriction tensor $Q_{ijkl}$ and the 4-th order rotostriction coefficients $R_{ijkl}$ are written in Vought notations. Forth order rotostriction coupling energy:

$$\begin{aligned}
G_4^{striction} &= -V_{111}\left(\sigma_1\Phi_1^4+\sigma_2\Phi_2^4+\sigma_3\Phi_3^4\right)-6V_{112}\left[\sigma_1\Phi_1^2\left(\Phi_2^2+\Phi_3^2\right)+\sigma_2\Phi_2^2\left(\Phi_3^2+\Phi_1^2\right)+\sigma_3\Phi_3^2\left(\Phi_1^2+\Phi_2^2\right)\right]\\
&-V_{122}\left[\sigma_1\left(\Phi_2^4+\Phi_3^4\right)+\sigma_2\left(\Phi_3^4+\Phi_1^4\right)+\sigma_3\left(\Phi_1^4+\Phi_2^4\right)\right]-6V_{123}\left[\sigma_1\Phi_2^2\Phi_3^2+\sigma_2\Phi_1^2\Phi_3^2+\sigma_3\Phi_2^2\Phi_1^2\right]\\
&-8V_{662}\left(\sigma_4\Phi_2\Phi_3\left(\Phi_2^2+\Phi_3^2\right)+\sigma_5\Phi_3\Phi_1\left(\Phi_3^2+\Phi_1^2\right)+\sigma_6\Phi_1\Phi_2\left(\Phi_1^2+\Phi_2^2\right)\right)\\
&-24V_{441}\left(\sigma_4\Phi_2\Phi_3\Phi_1^2+\sigma_5\Phi_3\Phi_1\Phi_2^2+\sigma_6\Phi_1\Phi_2\Phi_3^2\right)\\
&-W_{111}\left(\sigma_1\Psi_1^4+\sigma_2\Psi_2^4+\sigma_3\Psi_3^4\right)-6W_{112}\left[\sigma_1\Psi_1^2\left(\Psi_2^2+\Psi_3^2\right)+\sigma_2\Psi_2^2\left(\Psi_3^2+\Psi_1^2\right)+\sigma_3\Psi_3^2\left(\Psi_1^2+\Psi_2^2\right)\right]\\
&-W_{122}\left[\sigma_1\left(\Psi_2^4+\Psi_3^4\right)+\sigma_2\left(\Psi_3^4+\Psi_1^4\right)+\sigma_3\left(\Psi_1^4+\Psi_2^4\right)\right]-6W_{123}\left[\sigma_1\Psi_2^2\Psi_3^2+\sigma_2\Psi_1^2\Phi_3^2+\sigma_3\Phi_2^2\Phi_1^2\right]\\
&-8W_{662}\left(\sigma_4\Psi_2\Psi_3\left(\Psi_2^2+\Psi_3^2\right)+\sigma_5\Psi_3\Psi_1\left(\Psi_3^2+\Psi_1^2\right)+\sigma_6\Psi_1\Psi_2\left(\Psi_1^2+\Psi_2^2\right)\right)\\
&-24W_{441}\left(\sigma_4\Psi_2\Psi_3\Psi_1^2+\sigma_5\Psi_3\Psi_1\Psi_2^2+\sigma_6\Psi_1\Psi_2\Psi_3^2\right)
\end{aligned} \tag{C.5b}$$

The 6-th order rotostriction coefficients $V_{ijklmn}$ and $W_{ijklmn}$ are written in Vought notations.

Flexoelectric coupling energy:

$$f_{Flexo}=\frac{f_{ijkl}}{2}\left(u_{ij}\frac{\partial P_k}{\partial x_l}-P_k\frac{\partial u_{ij}}{\partial x_l}\right)\equiv\frac{F_{ijkl}}{2}\left(\sigma_{ij}\frac{\partial P_k}{\partial x_l}-P_k\frac{\partial \sigma_{ij}}{\partial x_l}\right) \tag{C.6}$$

$f_{ijkl}$ is the forth-rank tensor of flexoelectric coupling, $c_{ijkl}$ is the elastic stiffness.

Gradient energy is

$$f_{Gradient}=\frac{g_{ijkl}}{2}\left(\frac{\partial P_i}{\partial x_j}\frac{\partial P_k}{\partial x_l}\right)+\frac{v_{ijkl}}{2}\left(\frac{\partial \Phi_i}{\partial x_j}\frac{\partial \Phi_k}{\partial x_l}\right)+\frac{w_{ijkl}}{2}\left(\frac{\partial \Psi_i}{\partial x_j}\frac{\partial \Psi_k}{\partial x_l}\right). \tag{C.7}$$

The fourth-rank symmetric tensors of gradient energy, $g_{ijkl}$, $v_{ijkl}$ and $w_{ijkl}$ are positively defined. Note, that the symmetrical part of the matrix $(\partial P_i/\partial x_j)(\partial P_k/\partial x_l)$ contributes to the gradient energy of the bulk system [89].

Elastic strain energy is:

$$f_{Elastic}=\frac{c_{ijkl}(T,x)}{2}u_{ij}u_{kl}\rightarrow-\frac{s_{ijkl}(T,x)}{2}\sigma_{ij}\sigma_{kl}, \tag{C.8}$$

Summation is performed over all repeated indices.

Boundary conditions correspond to the vanishing of stresses far from the twin wall. SOP and polarization tends to their spontaneous constant values. Mechanical equilibrium equation $\partial\sigma_{ij}/\partial x_j=0$ should be valid.

$E_i^d=-\partial\varphi/\partial x_i$ are the components of electric (e.g. depolarization) field, caused by imperfect screening of the inhomogeneous polarization distribution with $\mathrm{div}(\mathbf{P})\neq 0$. The electrostatic potential, $\varphi$, satisfies the Poisson equation

$$\varepsilon_0\varepsilon_b\frac{\partial^2\varphi}{\partial x_i^2}=\frac{\partial P_i}{\partial x_i}-e\left(p(\varphi)+N_d^+-N_d^--n(\varphi)\right) \tag{C.9}$$

Here the charges are in the units of electron charge $e$=1.6×10$^{-19}$ C, $\varepsilon_0$=8.85×10$^{-12}$ F/m is the universal dielectric constant, $\varepsilon_b$ is the background dielectric permittivity of the material (unrelated with the soft mode), that is typically much smaller than the permittivity related with the soft mode. Boundary conditions to Eqs.(C.7) correspond to the electric potential vanishing far from the domain wall plane.

The density of free carriers will be considered in the simplest BPN approximation:

$$p(\varphi)=\int_0^\infty d\varepsilon\cdot g_p(\varepsilon)f(-\varepsilon-E_V+E_F+e\varphi)\approx p_0\exp\left(\frac{-\Xi_{ij}^V u_{ij}^S(x,z)-e\varphi}{k_B T}\right), \tag{C.10a}$$

$$n(\varphi)=\int_0^\infty d\varepsilon\cdot g_n(\varepsilon)f(\varepsilon+E_C-E_F-e\varphi)\approx n_0\exp\left(\frac{-\Xi_{ij}^C u_{ij}^S(x,z)+e\varphi}{k_B T}\right). \tag{C.10b}$$

Where $f(x)=\{1+\exp(x/k_BT)\}^{-1}$ is the Fermi-Dirac distribution function, $k_B$=1.3807×10$^{-23}$ J/K, $T$ is the absolute temperature. $E_F$ is the Fermi level, $E_C$ is the bottom of the conductive band, $E_V$ is the top of the valence band (all energies are defined with respect to the vacuum level). The equilibrium densities of holes and electrons are defined for the case φ = 0.

### C.2. Material parameters of $CaTiO_3$

In order to improve the situation we could use higher order roto-striction coupling, namely the spontaneous strain dependence on the order parameters could be written as

$$u_{ij}=R_{ijkl}\Phi_k\Phi_l+Z_{ijkl}\Psi_k\Psi_l+V_{ijklmn}\Phi_k\Phi_l\Phi_m\Phi_n+W_{ijklmn}\Psi_k\Psi_l\Psi_m\Psi_n \qquad \text{(C.11)}$$

Here the first two terms correspond to 2$^{nd}$ order rotostriction, while the latter two are for 4$^{th}$ order roto-striction. Note that we neglected the terms like $\Phi_k\Phi_l\Psi_m\Psi_n$ for clarity.

Considering high temperature aristotype **m3m** and taking into account internal symmetry of 4$^{th}$-order rotostriction, Eq.(C.11) can be essentially simplified. Namely, considering order parameter of for **4/mmm-phase** one could get from (C.11)

$$u_{11}=R_{12}\Phi_3^2+V_{122}\Phi_3^4,\quad u_{33}=R_{11}\Phi_3^2+V_{111}\Phi_3^4,\quad u_{12}=0. \qquad \text{(C.12)}$$

and for **mmm**-phase

$$u_{11}=R_{11}\Phi_1^2+R_{12}\Phi_2^2+Z_{12}\Psi_3^2+V_{111}\Phi_1^4+6V_{112}\Phi_1^2\Phi_2^2+V_{122}\Phi_2^4+W_{122}\Psi_3^4, \qquad \text{(C.13a)}$$

$$u_{33}=R_{12}\left(\Phi_1^2+\Phi_2^2\right)+Z_{11}\Psi_3^2+V_{122}\Phi_1^4+6V_{123}\Phi_1^2\Phi_2^2+V_{122}\Phi_2^4+W_{111}\Psi_3^4, \qquad \text{(C.13b)}$$

$$u_{12}=\frac{1}{2}R_{44}\Phi_1\Phi_2+4V_{662}\Phi_1\Phi_2\left(\Phi_1^2+\Phi_2^2\right). \qquad \text{(C.13c)}$$

Here we used Voigt matrix notations for 4$^{th}$ order rotostriction, i.e. $V_{ijklmn}\equiv V_{\alpha\beta\gamma}$ where one Greek index replaces two successive tensor indices.

We calculated pseudo-cubic lattice constants from the spontaneous strains according to equations

$$\frac{a}{\sqrt{2}}=a_0\left(1+u_{11}+u_{12}\right),\ \frac{b}{\sqrt{2}}=a_0\left(1+u_{11}-u_{12}\right)\qquad \frac{c}{2}=a_0\left(1+u_{33}\right), \qquad \text{(C.14)}$$

where $a$, $b$ and $c$ are the lattice parameters, and $a_0$ is the lattice parameter with cubic (m3m) symmetry. The lattice parameter $a_0$ is obtained from the extrapolation of a nonlinear fit with high temperature, and the formula is (see e.g. Ref.[90]):

$$a_0=a_{RT}+\alpha_T\Theta_S\left(\coth\left(\frac{\Theta_S}{T}\right)-\coth\left(\frac{\Theta_S}{T_{RT}}\right)\right) \qquad \text{(C.15)}$$

with a saturation temperature $\Theta_S$, room temperature lattice constant $a_{RT}$ =3.828 Å given by the extrapolation on the solid solution and high temperature expansion coefficient $\alpha_T$. Allowing for the discrepancy in the Kennedy [91] and Yashima [92] experimental data, parameters $\Theta_S$ and $\alpha_T$ fitted to Eq.(C.14) are different for different data sets (see **Table C1**). Corresponding temperature dependences of lattice parameters $a_0$ are shown in **Figure C1**.

**Table C1.** The fitted parameters for lattice parameter $a_0$

| Experimental data | $\alpha_T$ ($\times 10^{-5}$ Å/K) | $\Theta_S$ (K) |
|---|---|---|
| Kennedy | 6.289 | 640.0 |
| Yashima (fit 1) | 6.570 | 899.1 |
| Yashima (fit 2) | 5.91667 | 740 |

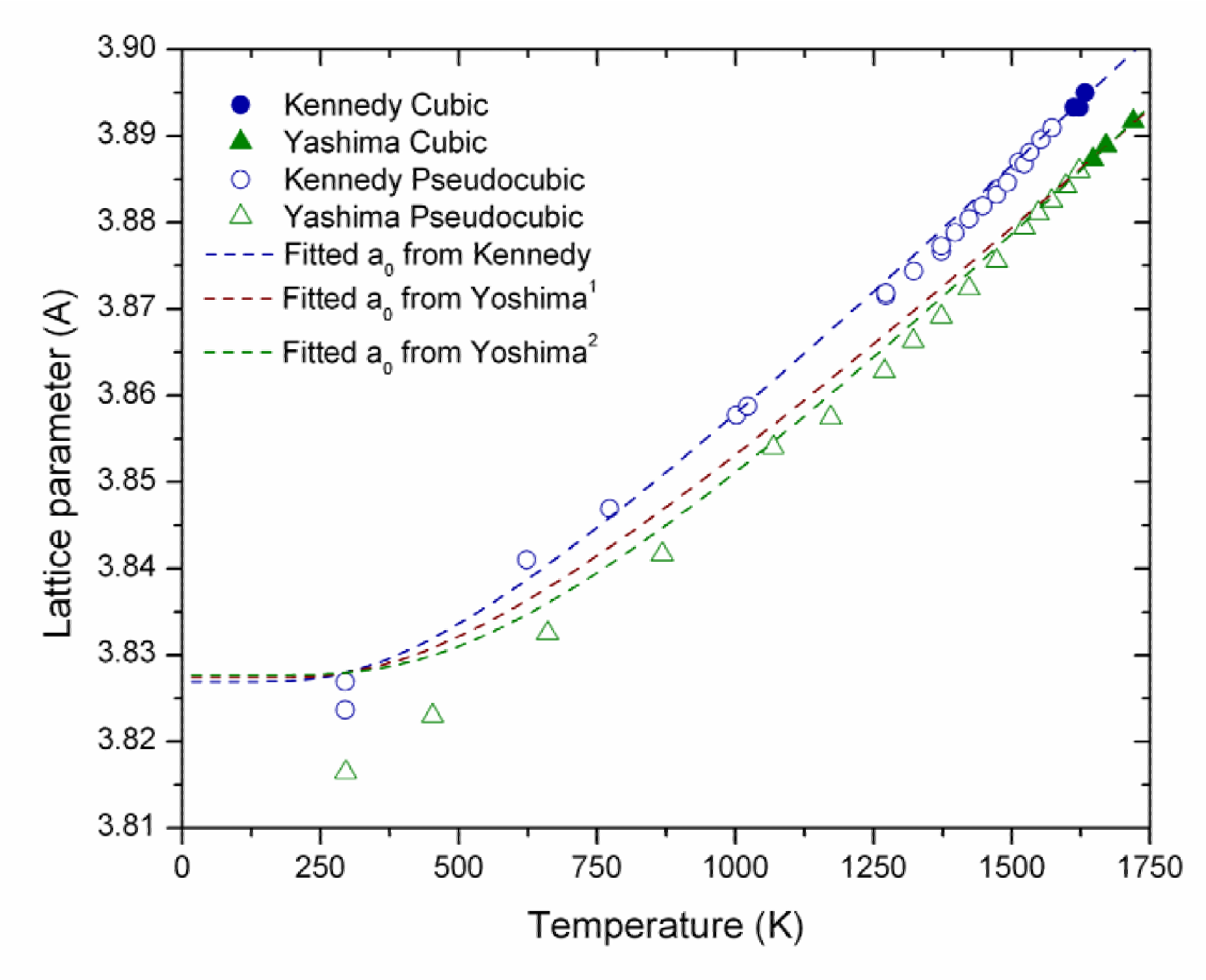

**Figure C1.** Temperature dependence of $CaTiO_3$ lattice parameter at m3m phase. The room temperature point for hypothetical m3m phase is obtained in Ref [90] by extrapolation on the solid solution composition. The "Pseudocubic" lattice constants are calculated from the cubic root of the volume, while the "Cubic" lattice parameters are from the cubic (m3m) phase.

Below we use only Yashima et al data for rotostriction coefficients fitting, since the authors measured both lattice constants in mmm, 4/mmm and m3m phases as well as corresponding spontaneous oxygen displacements for the same samples. Different fittings #1-3 are listed and described in the **Table C2**. Temperature dependences of $CaTiO_3$ pseudo-cubic lattice constants in mmm, 4/mmm and m3m3 phases and "cubic phase" (along with different fitting curves) are shown in **Fig.C2a.** Corresponding temperature dependences of the spontaneous oxygen displacements [93], which where used for the numerical fitting of the lattice constants, are shown in **Fig.C2b**. Temperature dependences of spontaneous strains in mmm and 4/mmm phases (points are recalculated from Ref. [93]) along with fitting (**set #1**) for temperature-dependent 2$^{nd}$ order rotostriction coefficients and fitting with the temperature-independent 2$^{nd}$ and 4$^{nd}$ order rotostriction coefficients (**set #2**) are shown in **Fig.C2c**. Fitting of experimental data with temperature-dependent 2$^{nd}$ order rotostriction and temperature-independent 4$^{nd}$ order rotostriction coefficients (**set #3)** is shown in **Fig.C2d**.

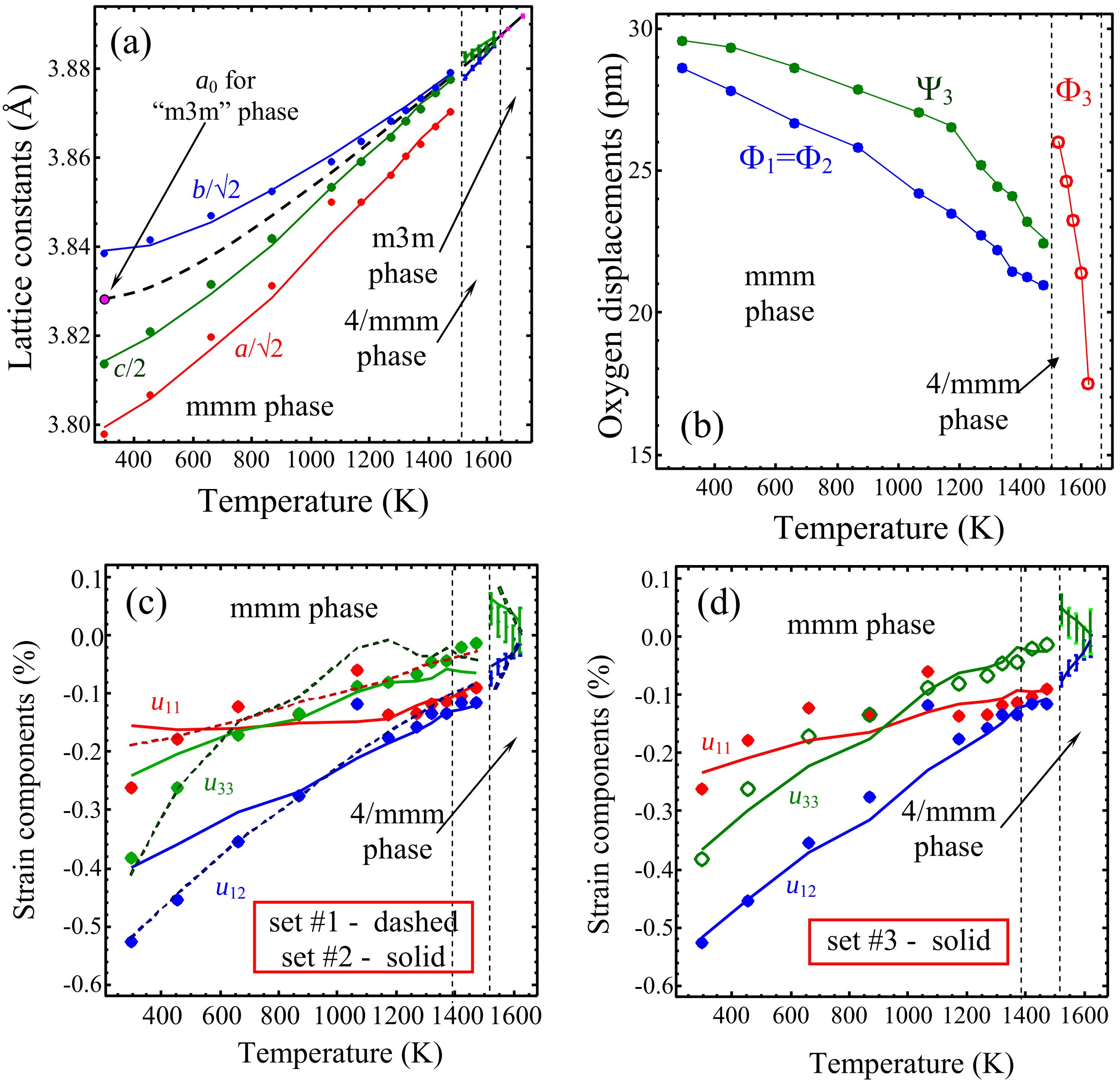

**Fig. C2**. (a) Temperature dependence of $CaTiO_3$ pseudo-cubic lattice constants in mmm, 4/mmm and m3m3 phases as well as "cubic phase" lattice constant (experimental data from Ref. [93], symbols) along with fitting curves (solid curves) (b) Temperature dependences of spontaneous oxygen displacements in mmm (solid dots +line) and 4/mmm (open dots + line) phases as recalculated from Ref.[93], which where used for the numerical fitting of the lattice constants (a). Room temperature point for hypothetical m3m phase is obtained in Ref. [94]. (c) Temperature dependence of spontaneous strains in mmm and 4/mmm phases (symbols are experimental data recalculated from Ref. [93]) along with fittings for the temperature-independent (solid curves, set #2) and temperature-dependent rotostriction coefficients (dashed curves, set #1). (d) Fitting Temperature-dependent with nonlinear striction (solid curves, set #3).

**Table C2**. Fitted roto-strictive coefficients

| **Fitting description, #** | **2nd order rotostriction coefficients ($\times 10^{17}$ m$^{-2}$)** | **4nd order rotostriction coefficients ($\times 10^{39}$ m$^{-4}$)** |
|---|---|---|
| **Set #1.** Temperature-dependent fitting without nonlinear striction | $R_{11}=r_{11}(T-T_{R11})=-0.154(T-1641.1)$, $R_{12}=r_{12}(T-T_{R12})=0.132(T-1648.6)$, $R_{44}=r_{44}(T-T_{R44})=0.076(T-1966.1)$, $Z_{11}=z_{12}(T-T_{Z11})=-0.217(T-1621.1)$, $Z_{12}=z_{12}(T-T_{Z12})=0.0345(T-1697.2)$ | $V_{ijk}=W_{ijk}\equiv 0$ The order parameters can be directly calculated from the potential [59] at each temperature |
| **Set #2**. Temperature-independent fitting with nonlinear striction | $R_{11}=6$, $R_{12}=-5$, $Z_{11}=8$, $Z_{12}=-14$, $R_{44}=-6$ | $V_{122}=-0.4$, $V_{111}=0.5$, $V_{112}=0.3$, $W_{122}=-2.2$, $V_{123}=-0.7$, $W_{111}=1.4$, $V_{662}=-0.7$ |
| **Set # 3.** Temperature-dependent fitting with nonlinear striction | $R_{11}=R_{o11}(1+r_{T11}T)=-3(1+10^{-4}T)$, $R_{12}=R_{o12}(1+r_{T12}T)=6(1-10^{-4}T)$, $R_{44}=R_{o44}(1+r_{T44}T)=15(1+6\cdot 10^{-4}T)$, $Z_{11}=Z_{o11}(1+z_{T11}T)=11(1+4\cdot 10^{-4}T)$, $Z_{12}=Z_{o12}(1+z_{T12}T)=-5(1+6\cdot 10^{-4}T)$ | $V_{122}=-2.4$, $V_{111}=1.6$, $V_{112}=-0.7$, $W_{122}=1.7$, $V_{123}=-0.9$, $W_{111}=1.5$, $V_{662}=-1.1$ |

It is seen from the figure that 2nd order rotostriction coefficients (**set #1**) are not enough to fit quantitatively the lattice parameter temperature dependence in different structural phases. In particular **set #1** overestimates the spontaneous strain in the tetragonal phase due to the high values of $R_{11}$ and $R_{12}$ with sharp temperature dependence. Fitting with the temperature-independent 2nd and 4nd order rotostriction coefficients (**set #2**) essentially underestimates the spontaneous strains at room temperature. Fitting with the temperature-independent 2nd order rotostriction coefficients and temperature-independent 4nd order rotostriction coefficients (**set #3**) looks the most adequate.

**Figures C3** demonstrate the values of the maximal strain $u_{xx}$ calculated using different sets of rotostriction coefficients. Plots a-c are calculated from Eqs.(C.12)-(C.13a) with the sets #1-3 correspondingly. It is seen from the figure that the using of "**set #1**" allowing for the temperature-dependent 2-nd order rotostriction, but without 4-th order rotostriction gives at least 2 times higher strain variation at the twin wall - surface junction that the **set #3**, at that the spontaneous strain far from the wall almost coincide for the **sets #1** and **#3**. Using of the **set #2** gives much smaller strain variation than the **sets #1** and **#3.** Consequently the improper polarization values calculated using the "**set #1**" appeared 2 times higher than the one for the **set #3** and 5 times higher than the one for the **set #2**. Carriers density values calculated using the "**set #1**" appeared an order of magnitude higher than for the **set #3**. Despite using of the **set #1** leads to highest improper polarization and carrier density accumulation at the twin walls, which is important for applications, **set #1** overestimates the spontaneous strain in the tetragonal phase due to the high values of $R_{11}$ and $R_{12}$ with sharp temperature dependence. Thus we decided to use **set #3** in the main paper (see **Table C3**).

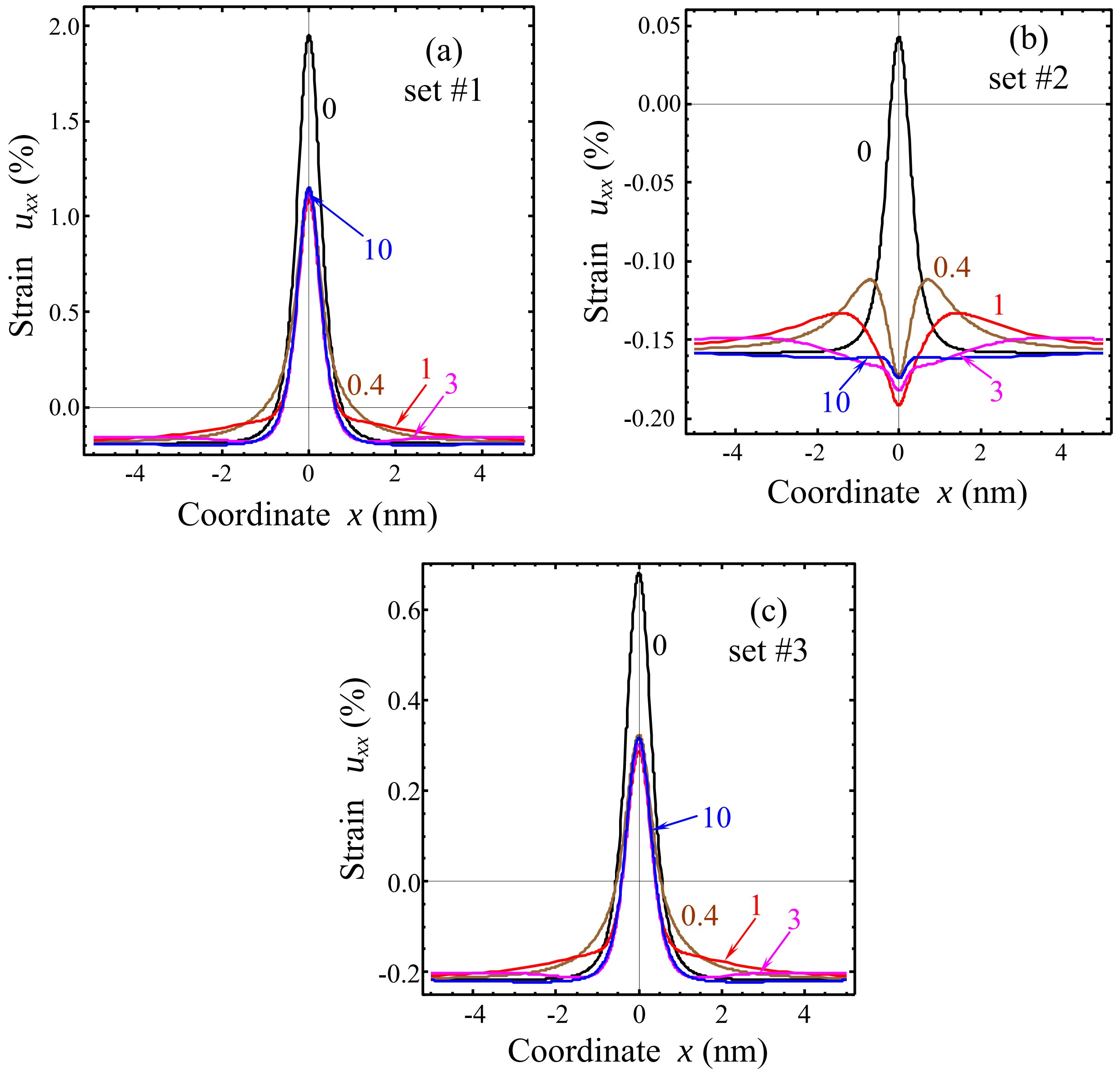

**Figures C3.** X-profiles of strain $u_{xx}$ calculated using different sets #1-3 of rotostriction coefficients. Plots a-c are calculated from Eqs.(C.12)-(C.13a) with the sets #1-3 correspondingly.

**Table C3.** $CaTiO_3$ material parameters at fixed stresses (recalculated mostly from Ref.[a])

| Parameter | SI units | Value | Source and notes |
|---|---|---|---|
| $\varepsilon_b$ | dimensionless | 58 | [b], experiment |
| $\alpha_T$ | $10^6\times$m/(F K) | 1.77 | [a], DFT |
| $\Theta_{s1}$ | K | 55 | [a], fitting |
| $T_1$ | K | 252 | [a], fitting |
| $a_{ij}$ | $m^5/(C^2F)$ | $\alpha_{11}= 2.90\times10^8$, $\alpha_{12}= 3.95\times10^6$ | [a], DFT |
| $Q_{ijkl}$ | $m^4/C^2$ | $Q_{11}$=0.0306, $Q_{12}$= -0.0099, $Q_{44}$=0.0771 | [a], DFT |
| $s_{ij}$ | $10^{-12}\times m^3/J$ | $s_{11}$=2.79, $s_{12}$= –0.59, $s_{44}$=10.01 | [a], DFT |
| $\beta_T$ | $10^{26}\times J/(m^5$ K) | 6.16 | [a], DFT |
| $\Theta_{s2}$ | K | 274 | [a], fitting |
| $T_2$ | K | 1590 | [a], fitting |
| $\beta_{ij}$ | $10^{49}\times J/m^7$ | $\beta_{11}$= - 2.256, $\beta_{12}$= - 5.774 | [a], fitting |
| $\beta_{ijk}$ | $10^{70}\times J/m^9$ | $\beta_{111}$=9.28, $\beta_{112}$=7.36 | [a], fitting |
| $\gamma_T$ | $10^{26}\times J/(m^5$ K) | 6.72 | [a], DFT |
| $\Theta_{s3}$ | K | 345 | [a], fitting |
| $T_3$ | K | 1285 | [a], fitting |
| $\gamma_{ij}$ | $10^{50}\times J/m^7$ | $\gamma_{11}+\gamma_{12}$= - 5.408 | [a], fitting |
| $\gamma_{ijk}$ | $10^{70}\times J/m^7$ | $\gamma_{111}+\gamma_{112}$=7.36 | [a], fitting |
| $\mu_{ij}$ | $10^{50}\times J/m^7$ | $\mu_{11}$ = -11.850, $\mu_{12}$ = 0.755; | [a], DFT |
| $t_{ij}$ | $10^{29}$ (F m)$^{-1}$ | $t_{11}$= -6.152, $t_{12}$= -3.848, $t_{44}$= -8.976 | [a], DFT |
| $\kappa_{ij}$ | $10^{29}$ (F m)$^{-1}$ | $\kappa_{11}$= -5.50, $\kappa_{12}$= -2.81 | [a], DFT |
| $R_{ij}$ | $10^{17}\times m^{-2}$ | $R_{11}=-3\left(1+10^{-4}T\right)$, $R_{12}=6\left(1-10^{-4}T\right)$, $R_{44}=15\left(1+6\cdot10^{-4}T\right)$. | fitting to spontaneous strain |
| $V_{ijk}$(**) | $10^{39}\times m^{-4}$ | $V_{111}$=1.6, $V_{112}$ = – 0.7, $V_{122}$ = – 2.4, $V_{123}$= – 0.9, $V_{662}$ = – 1.1 | |
| $Z_{ij}$ | $10^{17}\times m^{-2}$ | $Z_{11}=11\left(1+4\cdot10^{-4}T\right)$, $Z_{12}=-5\left(1+6\cdot10^{-4}T\right)$ | |
| $W_{ijk}$(**) | $10^{39}\times m^{-4}$ | $W_{111}$=1.5, $W_{122}$=1.7, | |
| $v_{ijkl}$ and $w_{ijkl}$ | $10^{10}\times J/m^3$ | Not used in the current study | unknown |
| $g_{ijkl}$ | $10^{-11}\times V\cdot m^3/C$ | Not used in the current study | unknown |
| $f_{ijkl}$ | V | $f_{11}$= 16 , $f_{12}$= -7 , $f_{44}$= 5. | Not measured |
| $F_{ijkl}$ | $10^{-11}\times m^3/C$ | $F_{11}$= 13.8, $F_{12}$= 6.66, $F_{44}$= 8.48 | Not measured |

[a] Yijia Gu, Karin Rabe, Eric Bousquet, Venkatraman Gopalan, and Long-Qing Chen. Phys. Rev. B **85**, 064117 (2012).

[b] G.Rupprecht & B.O. Bell. Phys. Rev. 135, A748 (1964).

*) The expansion coefficients (forth rank tensors) are at given stress (free sample), as recalculated from the ones at given strain (clamped sample)

**) Components of 6th rank tensor of 4th order roto-striction strain.